# Forecasting Labor Demand:
# Predicting JOLT Job Openings using Deep Learning Model

Kyungsu Kim[1]

[1] Georgia Institute of Technology School of Engineering, Atlanta GA 30332, USA

**Abstract.** This thesis studies the effectiveness of Long Short Term Memory (LSTM) model in forecasting future Job Openings and Labor Turnover Survey (JOLT) data in the United States. Drawing on multiple economic indicators from various sources, the data are fed directly into LSTM model to predict JOLT job openings in subsequent periods. The performance of the LSTM model is compared with conventional autoregressive approaches, including ARIMA, SARIMA, and Holt-Winters. Findings suggest that the LSTM model outperforms these traditional models in predicting JOLT job openings, as it not only captures the dependent variable's trends but also harmonized with key economic factors. These results highlight the potential of deep learning techniques in capturing complex temporal dependencies in economic data, offering valuable insights for policymakers and stakeholders in developing data-driven labor market strategies.

## 1    Introduction

### 1.1    Background of JOLT job opening data

This thesis focuses on JOLT Open data, a dataset derived from surveys such as the U.S. Bureau of Labor Statistics' Job Openings and Labor Turnover Survey (JOLTS). This data provides detailed insights into job openings, hires, and separations, offering a comprehensive view of labor market dynamics that extend beyond traditional unemployment statistics. It effectively captures the fluidity and evolving demand for labor, reflecting both employer intentions and worker movements within the economy.

Predicting JOLT data give us insights in various ways, below are some examples.
<u>Early Warning Signs:</u> Predicting job openings, hires, and separations can signal shifts in the economy, helping policymakers anticipate and respond to potential downturns before they escalate.
<u>Targeted Policy Adjustments:</u> With reliable forecasts, governments and central banks can fine-tune fiscal and monetary policies to better control inflation, boost economic growth, or cushion the impact of recessions.
<u>Efficient Resource Allocation:</u> Accurate labor market predictions help guide investments in workforce training and education, ensuring resources are directed where they are needed most.

### 1.2    Research Objectives and Motivations:

The primary objective of this research is to forecast economic impacts by predicting JOLT data. By accurately modeling job openings, hires, and separations, the study aims to provide insights into broader economic trends and facilitate proactive economic planning.

Initial experiments using traditional methods, such as AutoRegressive models, revealed limitations in capturing the complex, nonlinear patterns inherent in JOLT data. These shortcomings motivated the adoption of deep learning techniques, which offer enhanced modeling capabilities and improved predictive accuracy for intricate time-series data.

## 2 Literature Review

### 2.1 Conventional Autoregressive models and it's Limitations

There has been extensive research comparing the effectiveness of AR models and machine learning algorithms in predicting economic indices. While traditional time series models, such as AutoRegressive ,ARIMA, SARIMA have been widely used for macroeconomic forecasting, recent studies suggest that ML models can outperform them in certain cases. For instance, Premraj[1] found that for long time series with a rich set of variables, ML algorithms performed better than traditional TS regression models, particularly in the case of the USA.

For a particular long time series with a rich number of variables (e.g. USA), we see that the ML algorithms outperform the traditional TS regression models (PirasantPremraj, 2019) Since the number of job openings in the USA is influenced by various economic factors, it is essential to add the relevant macroeconomic indicators when modeling its trends. Even for the macroeconomic level, S&P500 index are one of the significant factors that affecting unemployment rate of the United States (Alrick Green,, 2024)[2]

In AutoRegressive (AR) models including ARIMA,SARIMA,Holt-Winters model, independent variables are not explicitly included as in standard regression models. Instead, AR models belong to the class of time series models, where the dependent variable is modeled based on its own past values.Also overall training method is focused on previous dependent variable, which make models vulnerable to external change of independent variables.And this is why most of the Machine Learning models and Deep Learning models are better than AutoRegressive based model when predicting economic indicators, because in Machine Learning based models we can add various external features to the considerations.

The AutoRegressive (AR) model follows a simple structure where the dependent variable is regressed on its own past values. AutoRegressive model's learning method is self-supervised training. The model generates a token based on the previously generated sequence [3]. If the self-generated token has bias, model itself keeps learning same bias over and over again for entire learning process.

$$X_t = \sum_{i=1}^{p} \varphi_i X_{t-i} + \varepsilon_t$$

Formula1. AutoRegressive model

The ARIMA (AutoRegressive Integrated Moving Average) model extends the AR model by incorporating differencing to ensure stationarity and adding a Moving Average (MA) component. Training an ARIMA model begins with model identification, where the data is checked for stationarity. If it's not stationary, differencing is applied to remove trends. The next step is determining stage where the best AR (p) and MA (q) terms using autocorrelation functions. Once the model structure is set, parameter estimation stage follows. This involves calculating the AR and MA coefficients using techniques like Maximum Likelihood Estimation (MLE) or Least Squares.
After estimating the parameters, the model moves to fitting stage, where it learns from past observations. The parameters are fine-tuned through iterative adjustments to minimize forecasting errors.
In the final step, forecasting stage, the model applies the learned parameters to predict future values. These predictions are based on previously generated data and past errors. As described above, ARIMA model doesn't consider external features for training nor predicting.

$$\left(1 - \sum_{i=1}^{p'} \alpha_i L^i\right) X_t = \left(1 + \sum_{i=1}^{q} \theta_i L^i\right) \varepsilon_t$$

Formula2. ARIMA model

The SARIMA (Seasonal ARIMA) model further enhances ARIMA by introducing seasonal components to account for periodic patterns. SARIMA builds on ARIMA by adding seasonality, making it good choice for time-series data with repeating patterns. The training stage follows a few key steps:

The first stage is model identification, where the data is checked for stationarity. If needed, both regular and seasonal differencing are applied to remove trends. Then, the best AR (p), MA (q), and differencing (d) terms are selected using autocorrelation functions. Seasonal parameters (P, D, Q, m) are also determined, with m representing the seasonal cycle length.

Next stage is parameter estimation, where the model calculates the AR, MA, and seasonal component coefficients using methods like Maximum Likelihood Estimation (MLE) or Least Squares.

Once the parameters are estimated, the model moves to fitting. Here, it learns from actual past data, adjusting its parameters to reduce forecasting errors. Unlike machine learning models, SARIMA doesn't rely on backpropagation but instead uses statistical optimization techniques.

The final stage is forecasting, where the trained model predicts future values based on past data and residual errors. These predictions are grounded in real historical trends rather than self-generated outputs, making SARIMA a reliable choice for structured time-series forecasting. Just like ARIMA modeling, SARIMA also doesn't consider external features, which makes it less-useful for predicting economic indexes.

$$\left(1 - \sum_{i=1}^{p} \phi_i L^i\right)(1-L^s)^D(1-L)^d X_t = \left(1 + \sum_{i=1}^{q} \theta_i L^i\right)\left(1 + \sum_{j=1}^{Q} \Theta_j L^{js}\right)\varepsilon_t$$

Formula3. SARIMA model

The Holt-Winters (Triple Exponential Smoothing) model trains itself by continuously refining its parameters to best fit past data, relying on three key components: level (baseline value), trend (rate of change), and seasonality (repeating patterns). It begins by estimating initial values for these components based on the first few observations. As new data points are added, the model updates its parameters using three smoothing equations: level smoothing adjusts the baseline value, trend smoothing accounts for changes over time, and seasonal smoothing captures repeating cycles. The smoothing factors α (level), β (trend), and γ (seasonality) determine how much weight recent data carries compared to older values. The model continuously fine-tunes these parameters using methods like Least Squares or Maximum Likelihood Estimation to minimize forecasting errors. Once trained, it predicts future values by projecting the level, trend, and seasonal effects forward, adjusting dynamically with each new data point rather than generating artificial inputs.

$$L_t = \alpha(X_t - S_{t-s}) + (1-\alpha)(L_{t-1} + B_{t-1})$$
$$B_t = \beta(L_t - L_{t-1}) + (1-\beta)B_{t-1}$$
$$S_t = \gamma(X_t - L_t) + (1-\gamma)S_{t-s}$$
$$F_{t+m} = (L_t + mB_t) + S_{t+m-s}$$

Formula4. Holt-Winters model

Thus, in this thesis, experiment will be conducted for AR based model versus Machine Learning based model(LSTM) for predicting JOLT job opening data.

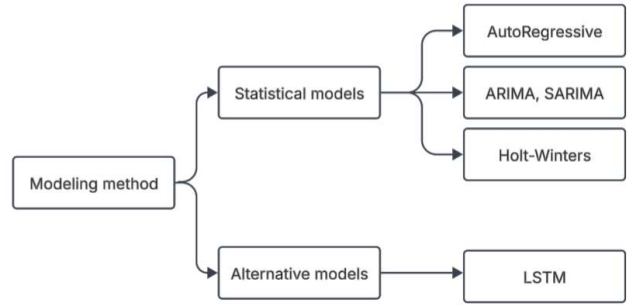

Figure1. Modeling method comparison covered in this thesis

## 2.2 Advantages of Deep Learning Model when predicting economic indicators

Due to inherent limitation of AR based models(AutoRegressive, ARIMA, SARIMA, Holt-Winters) performs lower when predicting multi-variate dependent values. AR models are simple and interpretable, they may struggle to capture the nonlinear dynamics present in many economic time series[6]. That is because AR based models are training based on it's self-generated tokens[2], which predict future values based on assumption that the future value of a variable depends linearly on its past values [6].

# 3 Methodology

## 3.1 Data Collection

The data used in this thesis were obtained from various reliable sources, including the U.S. Bureau of Labor Statistics (BLS), the U.S. Census Bureau, the U.S. Bureau of Economic Analysis (BEA), and the U.S. Energy Information Administration (EIA). These agencies provide publicly available economic data that are widely used in academic and policy research.

The data were accessed from the official databases of these institutions, ensures reliability and accuracy. The details of each indicator, including its definition and data source, are provided in Table 1.

**Table1**. Independent variable used in modeling and it's data source.

| Independent Variable Full Name | Data Source |
|---|---|
| United States Homeownership Rate | U.S. Census Bureau |
| United States Inflation Rate (Month-over-Month) | U.S. Bureau of Labor Statistics (BLS) |
| United States Producer Price Index (PPI) | U.S. Bureau of Labor Statistics (BLS) |
| United States Consumer Price Index (CPI) for Medical Care | U.S. Bureau of Labor Statistics (BLS) |
| United States Corporate Profits | U.S. Bureau of Economic Analysis (BEA) |
| United States Retail Sales Excluding Automobiles | U.S. Census Bureau |
| United States Current Account Balance | U.S. Bureau of Economic Analysis (BEA) |
| United States Personal Consumption Expenditures (PCE) | U.S. Bureau of Economic Analysis (BEA) |
| United States Foreign Direct Investment (FDI) | U.S. Bureau of Economic Analysis (BEA) |
| United States Import Price Index | U.S. Bureau of Labor Statistics (BLS) |
| United States Factory Orders | U.S. Census Bureau |
| United States Producer Price Index (PPI) for Commodities | U.S. Bureau of Labor Statistics (BLS) |
| United States Employment-to-Population Ratio | U.S. Bureau of Labor Statistics (BLS) |
| United States Personal Consumption Expenditures (PCE) Price Index | U.S. Bureau of Economic Analysis (BEA) |
| United States Crude Oil Runs to Refineries | U.S. Energy Information Administration (EIA) |
| United States Gasoline Stocks | U.S. Energy Information Administration (EIA) |
| United States Personal Savings Rate | U.S. Bureau of Economic Analysis (BEA) |
| United States Employment Payrolls Yearly | U.S. Bureau of Labor Statistics (BLS) |
| United States GDP from Industry | U.S. Bureau of Economic Analysis (BEA) |
| United States Economic Optimism Index | Economic Research Institutes |

To analyze the result, we need a deeper investigations.

United States Homeownership Rate, which represent percentage of U.S households that own their residences, could affect the job opening quantity in many ways. The more households purchase houses, we can think of it as U.S economic status is stronger. Stronger the economic status, more companies and government agents are tend to open more jobs, which makes JOLT job opening quantity increases.

United States Inflation Rate, which represent Consumer Price Index changes rate, also affects JOLT job openings. Companies and Public agents tends to minimize the uncertainty of its' operation. In 2022, after COVID-19 hits the labor market, job opening/vacancy affects overall inflation [8]. That's because higher the inflation, consumers spending is declining, causing industries decreasing production for goods and services, which lowers job openings.

United States Producer Price Index and United States Consumer Price Indexes which measuring the change of selling prices received by domestic producers and purchasing price of consumers, within the same context as United States Inflation Rate, are one of the main features affecting JOLT job openings.

United States Corporate Profits is indicator that describe total profits business profitability of U.S corporations. The national unemployment rate is expected to have a negative relationship with employment, median wages, and corporate profits [9].

United States Retail Sales Excluding Automobiles, measures the overall consumer spending on retail goods excluding vehicle sales. In various studies, increasing sales affecting new hiring, which results in higher JOLT job opening. Higher sales increase the likelihood of job creation significantly, The odds of hiring new employees increase by four times (4.0x) in firms that reported sales growth [10].

United States Current Account Balance, indicating the U.S transactions with the rest of the world. This index summarizes import and export amount, dividing sections into Primary Income (investment), Services, Secondary Income (Transfers), Goods. The key observation from these scenarios is the important synergies between growth, trade, and services, which is revealed by the fact that the overall improvements to the US current account trajectory from both higher growth through deeper integration of services and the new economy in domestic economies and increased income elasticities through increased trade in new economy services trade is greater than the either of these happening in alone [11].

United States Personal Consumption Expenditures, measures overall economic activity specifically how much does consumer spending on goods and services. In many studies and researches indicating that more the consumer spending on goods and services, PCE increases, the more jobs are created. Jobs that are created are the direct result of production in industries that produce goods and services to meet consumer demands (final goods), and the rest are generated in industries that provide inputs for the production of final goods and services (intermediate goods)]12].

United States Foreign Direct Investment, indicating foreign entities investment made to U.S corporations, also highly related to the JOLT job data. Because Foreign Direct Investments brings new businesses, factories it has direct relationship with JOLT job data. In particular, a 1% increase in manufacturing FDI leads to a 0.039% increase in manufacturing employment at the industry sector level in a given year. A 1% increase in GDP per capita leads to a 1.83% increase in manufacturing employment of an industry sector, of which the magnitude may reflect the potential for a large job growth that could follow economic growth in SSA countries [14]. A smaller technological or knowledge gap with the foreign firms is required for FDI to lead to more linkages and spillovers, and ultimately job creation for the poor. The results cast doubt on development strategies that rely on FDI as a sufficient policy for inclusive growth [14].

United States Import Price Index, indicating change in prices of goods imported into U.S also has high correlation with the JOLT data. Once import price rises, U.S manufacturers are facing increased cost of production, which result in low profitability. To respond that, manufacturers reducing hires or furthermore, layoff workers. If import price falls, conversely, U.S manufacturer's profitability increases, which could lead to hiring more employees. There has been several studies asserting that inverse relationship between import price and employment data [15], suggest that if the price of an imported good drops, people will buy fewer locally made substitutes, causing demand to fall and leading to job losses in those industries.

In a similar context, United States Factory Orders has similar impact on JOLT job openings. The more the factory orders, manufacturers are requiring more resources to produce such goods, which results in increasing hiring.

Moreover, factory orders are not just affecting manufacturing job openings, but also affecting overall supply chain and linked industries. For instance, decreased factory order cause less shipment from vendors, which also results in decreased demand for logistic services.

United States Producer Price Index (PPI) for Commodities, reflecting changes in selling price of commodities, also affecting labor demands. If this indicator increases, manufacturer's profit margin declines and may cut back on hiring and freeze job openings. If PPI gets lowered, the inflationary pressure declines and corporations can produce commodities with lower cost. This leads to higher profit margin, which leads to expanding operations, results in higher recruiting.

United States Employment-to-Population Ratio, which describe the proportion of U.S working-age population whom employed. High employment-to-population ratio indicates most people who can work, who wants to work are already working. This makes it harder for employer to find apt candidate for job and keep job openings open for longer period of time. Furthermore, when employment is high, fewer people left unemployed and actively looking for works, which results in lack of workforce. These shortages result in higher number of unfilled job openings and lead to higher job openings.

United States Personal Consumption Expenditures (PCE) Price Index, similar to PCE, measuring increase in prices with percentage not the amount. The impact on the job opening logic is similar.

United States Crude Oil Runs to Refineries, which measures volume of crude oil processed by refineries, if it's increasing, it means economic status is bright. Because better the economic status, the more transaction occurs, lead to more production and this results in higher oil demands. Various studies are indicating that the stronger the economic status, more jobs are created [16].

United States Gasoline Stocks, representing total gasoline inventories of United States also related to the job openings. When gasoline stocks are increasing it can indicate the supply of the gasoline is outpaced the demand, which results in lowering the price of gasoline. Once gasoline gets cheaper, it lowers the overall cost of the manufacturer such as overhead cost and logistic costs. This increase the profitability of producer, which results in higher job demand.

United States Personal Savings Rate, shows the percentage of how much of disposable income goes into savings. The U.S. Personal Savings Rate and job openings both shows how people and businesses treating the economy. When people feel confident about their jobs and economic status, they tends to spend more and save less. But if they're anxious about losing their job or the economy getting worse, they save more just in case. In the same way, a lot of job openings mean companies are doing well and hiring, while fewer job openings can mean businesses are being careful and the economy might be slowing down.

United States Employment Payrolls Yearly, which indicating changes in U.S payrolls. When a large number of jobs are added to the U.S. economy, it's often a sign that businesses are growing and looking to bring on more workers, which naturally leads to more job openings. For example, if companies add 300,000 jobs in a month, they probably had a lot of positions posted beforehand to support that growth. On the flip side, when job growth slows down or turns negative, employers might pause hiring, cut back on job postings, or even pull listings altogether, causing the number of job openings to drop. Generally, strong job growth shows that businesses feel confident about the economy and are ready to expand. But when there's uncertainty—like fears of a recession—companies often start holding back on hiring, sometimes even before unemployment numbers begin to reflect the slowdown.

United States GDP from Industry indicating GDP contributions from industries. An increase in GDP from the industrial sector, encompassing manufacturing, construction, and mining, typically signifies a rise in industrial output and overall economic productivity. This expansion necessitates a larger workforce to sustain production, thereby generating additional job openings across a range of occupational levels. Elevated industrial activity not only creates demand for specialized roles such as machine operators, engineers, and logistics personnel, but also stimulates employment in associated white-collar positions. Studies are suggesting that increases in GDP can increase employment due to unemployed people becoming employed and through a shift from the informal sector to the formal sector which does not increase employment but increases wages[18]. The growth of industrial output also produces a ripple effect across interconnected sectors—including transportation, warehousing, information technology, and various professional services—leading to a broader increase in job opportunities throughout the economy.

United States Economic Optimism Index, measuring how confidence and optimistic does economic outlook will be in near future, is also plays critical role in predicting job openings.

The U.S. Economic Optimism Index influences job openings by signaling the overall confidence of consumers and businesses in the economic outlook. When optimism increases, firms are more likely to anticipate growth in demand, prompting them to expand operations and recruit additional labor. In contrast, diminished confidence often leads to hiring freezes or workforce reductions as a precaution against economic uncertainty. Consumer sentiment plays a pivotal role, as higher confidence tends to stimulate spending, which in turn drives business revenue and labor demand[18].

### 3.2 Data Processing and Model architecture

Data processing starts with imputation. For objective comparison, I used one of the simplest methods, the backward fill (bfill) method, for imputing missing values. Then I applied the pre-processed dataset into tree-based CatBoost model. For feature extraction using the CatBoost model, the following aspects were not considered to ensure fair and objective comparison with other models: lag features, time series structure, and seasonality.
Subsequently, the top 20 features, which has the most influences upon model were extracted based on feature importance. These selected features were then applied to the AR, ARIMA, SARIMA, Holt-Winters, and LSTM models.

Figure2. AR model prediction and test data
AR model prediction shows the worst performance of all AR based models.

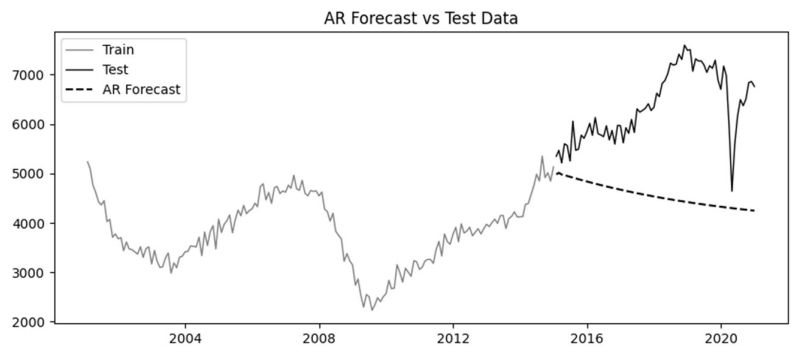

Figure3. ARIMA model prediction and test data
ARIMA model shows slight better performance than AR model, but still lack of captures the fluctuations.

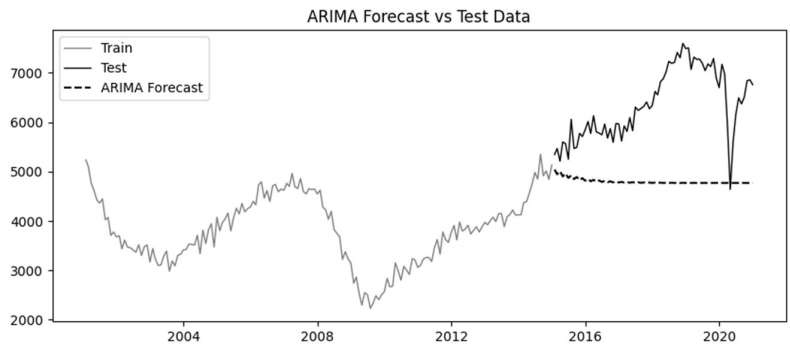

Figure4. SARIMA model prediction and test data
Since SARIMA considers seasonality, it shows slight fluctuations between months, but still doesn't captures the major fluctuations.

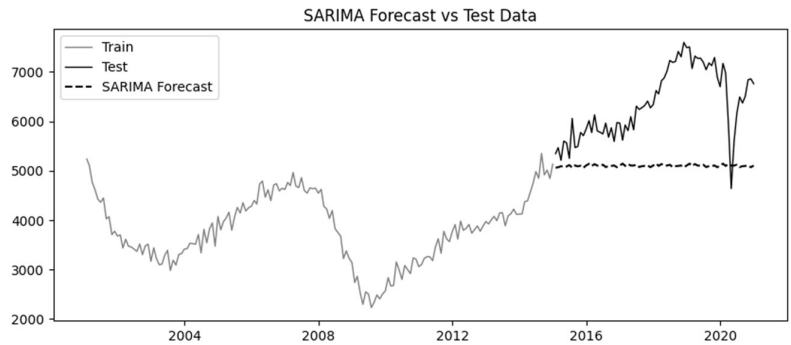

Figure5. Holt-Winters model prediction and test data
Because Holt-Winters breaks down time-series data into 3 part (level / trend / seasonality), it captures increasing trend of job increasing well, but lacks prediction power when it comes to decreasing trend.

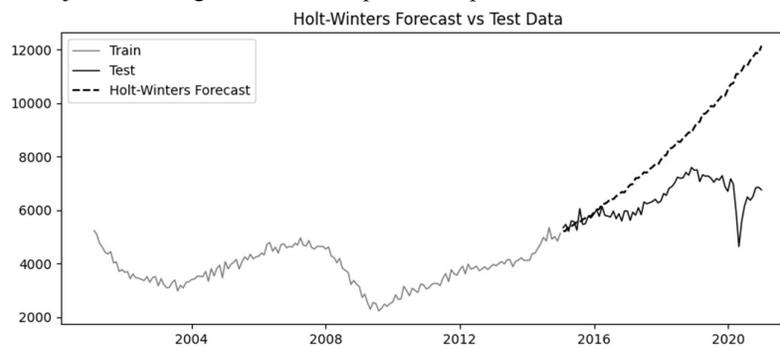

Figure6. LSTM model prediction and test data
Because LSTM model considers multivariate input and processing time-series data recurrently, it captures the fluctuation of the data and thus has better prediction power than AR based model.

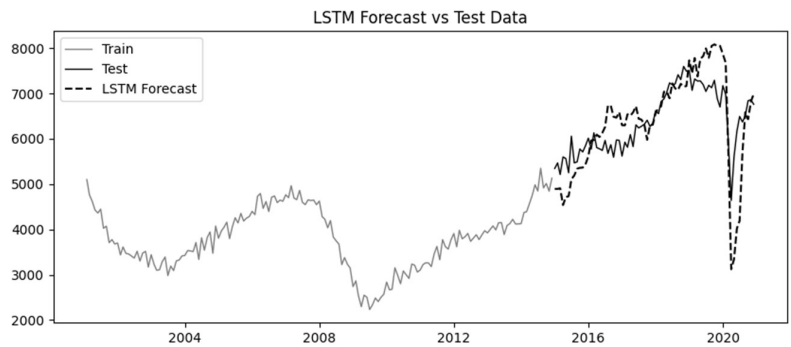

All models were built using monthly data from January 2001 to December 2020, with a time lag of 3. The training dataset consisted of the initial 70% of the data starting from January 2001, while the remaining 30% was used for testing and prediction.

In this study, I employed a deep Long Short-Term Memory (LSTM) network to predict target values(JOLT job openings) from multivariate time series data. LSTM was selected over traditional AR-based models due to its prediction performance, particularly in capturing complex, nonlinear temporal dependencies.

The input consists of a 20-dimensional feature vector representing multivariate time series data, which was selected by feature importance ranking of CatBoost model. All features were normalized using the median and the interquartile range (IQR), rather than the mean and standard deviation. This Robust Scaler was applied because economic data has high number of outliers, including COVID period. This approach makes it robust to outliers, as it reduces the influence of extreme values during the scaling process. Before being fed into the model. The proposed model architecture comprises six LSTM layers with the following number of units: 256, 64, 32, 32, 16, and 1. Each LSTM layer uses the hyperbolic tangent (tanh) activation function, followed by a linear transformation in the final output layer to accommodate the regression task. Reason why I chose mixing linear function and hyperbolic tangent function is because it performs the best. Regularization techniques such as dropout or weight normalization were intentionally excluded, as initial experiments indicated that their application degraded performance due to excessive decaying in temporal memory retention.

The model was trained using the Adam optimizer with the following hyperparameters: learning rate = 0.003, $\beta_1$ = 0.9, $\beta_2$ = 0.999, and $\varepsilon$ = 1e-7. The loss function used was Mean Squared Error (MSE). Training was conducted over 50 epochs with a batch size of 32, using a time step of 2 to capture short-term temporal patterns.

Figure7. Model architecture in forms of Neural Network.
It starts with 256 layer and then ends with predicting 1 dependent variable.
Height represent period(12), Width represent the number of features, Depth represent year(2001-2020)
Starts with 318 features of 12 period with 20 years and then ends with 1 vector space, which predicts JOLT job openings for specific period.

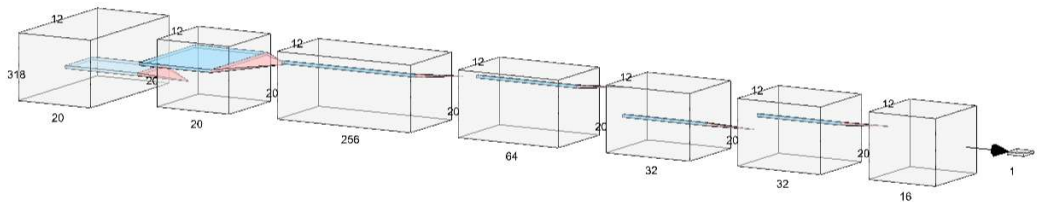

## 4  Results and Discussion

The research was conducted 50 iterations, and the average result of the prediction was measured to ensure the stability of the test. In this research, experiment was conducted to predict JOLTS(Job Openings and Labor Turnover Survey) job opening amount, which represent the current status of labor market. The model used for prediction was AR(Autoregressive), (Autoregressive Integrated Moving Average), SARIMA (Seasonal ARIMA), Holt-Winters Exponential Average,and LSTM (Long Short-Term Memory) Neural Network model.

For comparison, RMSE(Root Mean Square Error) and MAPE(Mean Absolute Error Percentage) was used for metric. Among the model, LSTM demonstrated the best predictive performance, showing the lowest error scores. This is because LSTM model is multivariate based model, which takes the number of economic exogenous variables as input. This capability enables LSTM model to capture the complex non-linearity aspects of JOLT data and enables LSTM to performs better than traditional time-series models.

These results suggest that time series based data characterized by nonlinearity and long-term dependencies, such as economic indicators, deep learning-based approaches like LSTM performs significant advantages in forecasting accuracy.

Table 2. RMSE, MAPE metric of Predicting JOLT Job openings per each Lag parameter

| Lag period | Root Mean Squared Error | | | | | Mean Absolute Percentage Error | | | | |
| --- | --- | --- | --- | --- | --- | --- | --- | --- | --- | --- |
| | 1 | 2 | 3 | 4 | Mean | 1 | 2 | 3 | 4 | Mean |
| AR | 2405.98 | 2418.13 | 1992.28 | 2047.58 | **2,215.99** | 33.51 | 29.44 | 27.11 | 27.86 | **29.48** |
| ARIMA | 1788.78 | 1792.32 | 1720.9 | 1784.38 | **1,771.60** | 24.72 | 24.78 | 23.63 | 24.65 | **24.45** |
| SARIMA | 1436.68 | 1404.82 | 1014.54 | 1455.28 | **1,327.83** | 19.11 | 18.56 | 12.82 | 19.01 | **17.38** |
| Holt-Winters | N/A* | 820.61 | 788.43 | 1429.78 | **1,012.94** | N/A* | 8.09 | 7.76 | 15.42 | **10.42** |
| LSTM | 785.41 | 740.18 | 755.38 | 1036.727 | **829.42** | 0.11 | 0.09 | 0.11 | 0.15 | **0.11** |

*Holt-Winters model with lag=1 is too short to capture the seasonality, so excluded from the experiment.

## Appendix A. Data Description

| Independent Variable | Description of indicator | Data Source |
| --- | --- | --- |
| UNITEDSTAHOMOWNRAT | United States Homeownership Rate | U.S. Census Bureau |
| UNITEDSTAINFRATMOM | United States Inflation Rate (Month-over-Month) | U.S. Bureau of Labor Statistics (BLS) |
| USAPPIM | United States Producer Price Index (PPI) | U.S. Bureau of Labor Statistics (BLS) |
| UNITEDSTACPIMED | United States Consumer Price Index (CPI) for Medical Care | U.S. Bureau of Labor Statistics (BLS) |
| UNITEDSTACORPROPRI | United States Corporate Profits | U.S. Bureau of Economic Analysis (BEA) |
| UNITEDSTARETSALEXAUT | United States Retail Sales Excluding Automobiles | U.S. Census Bureau |
| USACSA | United States Current Account Balance | U.S. Bureau of Economic Analysis (BEA) |
| USAPREC | United States Personal Consumption Expenditures (PCE) | U.S. Bureau of Economic Analysis (BEA) |
| UNITEDSTAFORDIRINV | United States Foreign Direct Investment (FDI) | U.S. Bureau of Economic Analysis (BEA) |
| UNITEDSTAIMPPRI | United States Import Price Index | U.S. Bureau of Labor Statistics (BLS) |
| USAFCAS | United States Factory Orders | U.S. Census Bureau |
| USAPPIMC | United States Producer Price Index (PPI) for Commodities | U.S. Bureau of Labor Statistics (BLS) |
| UNITEDSTAEMPPER | United States Employment-to-Population Ratio | U.S. Bureau of Labor Statistics (BLS) |
| USAPEFEATSM | United States Personal Consumption Expenditures (PCE) Price Index | U.S. Bureau of Economic Analysis (BEA) |
| UNITEDSTAAPICRURUN | United States Crude Oil Runs to Refineries | U.S. Energy Information Administration (EIA) |
| UNITEDSTAAPIGASSTO | United States Gasoline Stocks | U.S. Energy Information Administration (EIA) |
| UNITEDSTAPERSPE | United States Personal Savings Rate | U.S. Bureau of Economic Analysis (BEA) |
| USAEPY | United States Employment Payrolls Yearly | U.S. Bureau of Labor Statistics (BLS) |
| UNITEDSTAGDPFROMIN | United States GDP from Industry | U.S. Bureau of Economic Analysis (BEA) |

| date | UNITEDSTAHOMOWNRAT | UNITEDSTAINFRATMOM | USAPPIM | UNITEDSTACPIMED | UNITEDSTACORPROPRI | UNITEDSTARETSALEXAUT | USACSA | USAPREC | UNITEDSTAFORDIRINV | UNITEDSTAIMPPRI |
|---|---|---|---|---|---|---|---|---|---|---|
| 1/31/2001 | 67.5 | 0.6 | 0.1 | 3 | 99.9 | 0.3 | 175.6 | 685.54 | 8327 | 100.5 |
| 2/28/2001 | 67.5 | 0.2 | 0.1 | 3.1 | 99.9 | -0.2 | 176 | 685.54 | 8327 | 99.9 |
| 3/31/2001 | 67.5 | 0.1 | 0.1 | 3 | 99.9 | -0.8 | 176.1 | 685.54 | 5149 | 98.3 |
| 4/30/2001 | 67.5 | 0.2 | 0.1 | 3.1 | 99.9 | 1.5 | 176.4 | 685.54 | 5149 | 97.8 |
| 5/31/2001 | 67.5 | 0.5 | 0.1 | 3.1 | 99.9 | 0.3 | 177.3 | 685.54 | 5149 | 98 |
| 6/30/2001 | 67.7 | 0.2 | 0.1 | 3.2 | 99.9 | -0.4 | 177.7 | 685.54 | 3227 | 97.6 |
| 7/31/2001 | 67.7 | -0.2 | 0.1 | 3.1 | 99.9 | -0.1 | 177.4 | 685.54 | 3227 | 96.1 |
| 8/31/2001 | 67.7 | 0 | 0.1 | 3.2 | 99.9 | 0.8 | 177.4 | 685.54 | 3227 | 96 |
| 9/30/2001 | 68.1 | 0.4 | 0.1 | 3.1 | 99.9 | -1.4 | 178.1 | 685.54 | 5530 | 95.9 |
| 10/31/2001 | 68.1 | -0.3 | 0.1 | 3.3 | 99.9 | 0.9 | 177.6 | 685.54 | 5530 | 93.7 |
| 11/30/2001 | 68.1 | -0.1 | 0.1 | 3.3 | 99.9 | 0.3 | 177.5 | 685.54 | 5530 | 92.3 |
| 12/31/2001 | 68 | -0.1 | 0.1 | 3.3 | 99.9 | 0.4 | 177.4 | 697.12 | -9988 | 91.4 |
| 1/31/2002 | 68 | 0.2 | 0.1 | 3.2 | 99.9 | 0.2 | 177.7 | 697.12 | -9988 | 91.6 |
| 2/28/2002 | 68 | 0.2 | 0.1 | 3.2 | 99.9 | 0.5 | 178 | 697.12 | -9988 | 91.6 |
| 3/31/2002 | 67.8 | 0.3 | 0.1 | 3.2 | 99.9 | 0.4 | 178.5 | 697.12 | 5347 | 92.8 |
| 4/30/2002 | 67.8 | 0.4 | 0.1 | 3.2 | 99.9 | 1.3 | 179.3 | 697.12 | 5347 | 94.3 |
| 5/31/2002 | 67.8 | 0.1 | 0.1 | 3.1 | 99.9 | -0.3 | 179.5 | 697.12 | 5347 | 94.4 |
| 6/30/2002 | 67.6 | 0.1 | 0.1 | 3 | 99.9 | 0.1 | 179.6 | 697.12 | 11591 | 94.1 |
| 7/31/2002 | 67.6 | 0.2 | 0.1 | 3 | 99.9 | 0.2 | 180 | 697.12 | 11591 | 94.5 |
| 8/31/2002 | 67.6 | 0.3 | 0.1 | 2.8 | 99.9 | 0 | 180.5 | 697.12 | 11591 | 94.8 |
| 9/30/2002 | 68 | 0.2 | 0.1 | 2.8 | 99.9 | 0 | 180.8 | 697.12 | 11862 | 95.5 |
| 10/31/2002 | 68 | 0.2 | 0.1 | 2.7 | 99.9 | 0.8 | 181.2 | 697.12 | 11862 | 95.5 |
| 11/30/2002 | 68 | 0.2 | 0.1 | 2.5 | 99.9 | 0.7 | 181.5 | 697.12 | 11862 | 94.6 |
| 12/31/2002 | 68.3 | 0.2 | 0.1 | 2.6 | 99.9 | 0.3 | 181.8 | 692.23 | 3498 | 95.2 |
| 1/31/2003 | 68.3 | 0.4 | 0.1 | 2.5 | 99.9 | 0.6 | 182.6 | 692.23 | 3498 | 96.9 |
| 2/28/2003 | 68.3 | 0.5 | 0.1 | 2.4 | 99.9 | -0.4 | 183.6 | 692.23 | 3498 | 98.5 |
| 3/31/2003 | 68 | 0.2 | 0.1 | 2.3 | 99.9 | 1.4 | 183.9 | 692.23 | 14840 | 99.1 |
| 4/30/2003 | 68 | -0.4 | 0.1 | 2.1 | 99.9 | -0.7 | 183.2 | 692.23 | 14840 | 96 |
| 5/31/2003 | 68 | -0.2 | 0.1 | 2 | 99.9 | 0.3 | 182.9 | 692.23 | 14840 | 95.3 |
| 6/30/2003 | 68 | 0.1 | 0.1 | 1.9 | 99.9 | 1.1 | 183.1 | 692.23 | 14578 | 96.2 |
| 7/31/2003 | 68 | 0.3 | 0.1 | 1.9 | 99.9 | 1 | 183.7 | 692.23 | 14578 | 96.7 |
| 8/31/2003 | 68 | 0.4 | 0.1 | 1.9 | 99.9 | 1.4 | 184.5 | 692.23 | 14578 | 96.7 |
| 9/30/2003 | 68.4 | 0.3 | 0.1 | 1.9 | 99.9 | -0.1 | 185.1 | 692.23 | 16797 | 96.2 |
| 10/31/2003 | 68.4 | -0.1 | 0.1 | 1.8 | 99.9 | 0.1 | 184.9 | 692.23 | 16797 | 96.3 |
| 11/30/2003 | 68.4 | 0.1 | 0.1 | 1.8 | 99.9 | 0.8 | 185 | 692.23 | 16797 | 96.8 |
| 12/31/2003 | 68.6 | 0.3 | 0.1 | 1.7 | 99.9 | 0.2 | 185.5 | 722.32 | 14749 | 97.5 |
| 1/31/2004 | 68.6 | 0.4 | 0.1 | 1.7 | 99.9 | 1.1 | 186.3 | 722.32 | 14749 | 99 |
| 2/29/2004 | 68.6 | 0.2 | 0.1 | 1.8 | 99.9 | 0.1 | 186.7 | 722.32 | 14749 | 99.4 |
| 3/31/2004 | 68.6 | 0.2 | 0.1 | 1.9 | 99.9 | 1.9 | 187.1 | 722.32 | 17553 | 100.2 |
| 4/30/2004 | 68.6 | 0.2 | 0.1 | 2.1 | 99.9 | -0.5 | 187.4 | 722.32 | 17553 | 100.4 |
| 5/31/2004 | 68.6 | 0.4 | 0.1 | 2.2 | 99.9 | 1.1 | 188.2 | 722.32 | 17553 | 101.9 |
| 6/30/2004 | 69.2 | 0.4 | 0.1 | 2.4 | 99.9 | -0.1 | 188.9 | 722.32 | 25186 | 101.7 |
| 7/31/2004 | 69.2 | 0.1 | 0.1 | 2.3 | 99.9 | 0.5 | 189.1 | 722.32 | 25186 | 102.1 |
| 8/31/2004 | 69.2 | 0.1 | 0.1 | 2.3 | 99.9 | 0.1 | 189.2 | 722.32 | 25186 | 103.6 |
| 9/30/2004 | 69 | 0.3 | 0.1 | 2.4 | 99.9 | 0.9 | 189.8 | 722.32 | 22748 | 104.1 |
| 10/31/2004 | 69 | 0.5 | 0.1 | 2.4 | 99.9 | 1.2 | 190.8 | 722.32 | 22748 | 105.8 |
| 11/30/2004 | 69 | 0.5 | 0.1 | 2.4 | 99.9 | 0.6 | 191.7 | 722.32 | 22748 | 105.5 |
| 12/31/2004 | 69.2 | 0 | 0.1 | 2.4 | 99.9 | 0.8 | 191.7 | 772.9 | 22403 | 104 |
| 1/31/2005 | 69.2 | -0.1 | 0.1 | 2.5 | 99.9 | 0 | 191.6 | 772.9 | 22403 | 104.6 |
| 2/28/2005 | 69.2 | 0.4 | 0.1 | 2.6 | 99.9 | 1 | 192.4 | 772.9 | 22403 | 105.5 |
| 3/31/2005 | 69.1 | 0.4 | 0.1 | 2.5 | 99.9 | 0.1 | 193.1 | 772.9 | 25378 | 107.8 |
| 4/30/2005 | 69.1 | 0.3 | 0.1 | 2.5 | 99.9 | 1.3 | 193.7 | 772.9 | 25378 | 108.8 |

| Date | | | | | | | | | | |
|---|---|---|---|---|---|---|---|---|---|---|
| 5/31/2005 | 69.1 | -0.1 | 0.1 | 2.5 | 99.9 | -0.6 | 193.6 | 772.9 | 25378 | 107.9 |
| 6/30/2005 | 68.6 | 0.1 | 0.1 | 2.4 | 99.9 | 1.2 | 193.7 | 772.9 | 28241 | 109.2 |
| 7/31/2005 | 68.6 | 0.6 | 0.1 | 2.5 | 99.9 | 0.2 | 194.9 | 772.9 | 28241 | 110.5 |
| 8/31/2005 | 68.6 | 0.6 | 0.1 | 2.4 | 99.9 | 1.7 | 196.1 | 772.9 | 28241 | 112.1 |
| 9/30/2005 | 68.8 | 1.4 | 0.1 | 2.6 | 99.9 | 1.4 | 198.8 | 772.9 | 24148 | 114.4 |
| 10/31/2005 | 68.8 | 0.2 | 0.1 | 2.6 | 99.9 | 1.1 | 199.1 | 772.9 | 24148 | 114.5 |
| 11/30/2005 | 68.8 | -0.5 | 0.1 | 2.7 | 99.9 | -0.8 | 198.1 | 772.9 | 24148 | 112.3 |
| 12/31/2005 | 69 | 0 | 0.1 | 2.8 | 99.9 | 0 | 198.1 | 718.87 | 32558 | 112.3 |
| 1/31/2006 | 69 | 0.6 | 0.1 | 2.7 | 99.9 | 2.4 | 199.3 | 718.87 | 32558 | 113.7 |
| 2/28/2006 | 69 | 0.1 | 0.1 | 2.6 | 99.9 | 0 | 199.4 | 718.87 | 32558 | 112.8 |
| 3/31/2006 | 68.5 | 0.2 | 0.1 | 2.7 | 99.9 | 0.2 | 199.7 | 718.87 | 31339 | 112.7 |
| 4/30/2006 | 68.5 | 0.5 | 0.1 | 2.8 | 99.9 | 0.5 | 200.7 | 718.87 | 31339 | 115.1 |
| 5/31/2006 | 68.5 | 0.3 | 0.1 | 2.9 | 99.9 | 0 | 201.3 | 718.87 | 31339 | 117.2 |
| 6/30/2006 | 68.7 | 0.2 | 0.1 | 3 | 99.9 | 0 | 201.8 | 718.87 | 37817 | 117.3 |
| 7/31/2006 | 68.7 | 0.5 | 0.1 | 3 | 99.9 | 0.2 | 202.9 | 718.87 | 37817 | 118.2 |
| 8/31/2006 | 68.7 | 0.4 | 0.1 | 3.2 | 99.9 | 0.9 | 203.8 | 718.87 | 37817 | 118.8 |
| 9/30/2006 | 69 | -0.5 | 0.1 | 3 | 99.9 | -0.4 | 202.8 | 718.87 | 40327 | 116.2 |
| 10/31/2006 | 69 | -0.4 | 0.1 | 3.1 | 99.9 | -0.5 | 201.9 | 718.87 | 40327 | 113.3 |
| 11/30/2006 | 69 | 0 | 0.1 | 3.1 | 99.9 | 0.1 | 202 | 718.87 | 40327 | 113.8 |
| 12/31/2006 | 68.9 | 0.5 | 0.1 | 3.1 | 99.9 | 1.8 | 203.1 | 707.61 | 35272 | 115.1 |
| 1/31/2007 | 68.9 | 0.2 | 0.1 | 3.1 | 99.9 | 0 | 203.437 | 707.61 | 35272 | 113.7 |
| 2/28/2007 | 68.9 | 0.4 | 0.1 | 3.3 | 99.9 | 0 | 204.226 | 707.61 | 35272 | 114.1 |
| 3/31/2007 | 68.4 | 0.5 | 0.1 | 3.3 | 99.9 | 1.2 | 205.288 | 707.61 | 33833 | 115.9 |
| 4/30/2007 | 68.4 | 0.3 | 0.1 | 3.3 | 99.9 | -0.6 | 205.904 | 707.61 | 33833 | 117.5 |
| 5/31/2007 | 68.4 | 0.4 | 0.1 | 3.1 | 99.9 | 1.3 | 206.755 | 707.61 | 33833 | 118.6 |
| 6/30/2007 | 68.2 | 0.2 | 0.1 | 3.1 | 99.9 | -0.1 | 207.234 | 707.61 | 37327 | 120 |
| 7/31/2007 | 68.2 | 0.2 | 0.1 | 3.1 | 99.9 | 0.5 | 207.603 | 707.61 | 37327 | 121.5 |
| 8/31/2007 | 68.2 | 0 | 0.1 | 2.9 | 99.9 | 0 | 207.667 | 707.61 | 37327 | 121.1 |
| 9/30/2007 | 68.2 | 0.4 | 0.1 | 3 | 99.9 | 0.3 | 208.547 | 707.61 | 29654 | 121.8 |
| 10/31/2007 | 68.2 | 0.3 | 0.1 | 3 | 99.9 | 0.7 | 209.19 | 707.61 | 29654 | 123.6 |
| 11/30/2007 | 68.2 | 0.8 | 0.1 | 3 | 99.9 | 1.8 | 210.834 | 707.61 | 29654 | 127.5 |
| 12/31/2007 | 67.8 | 0.3 | 0.1 | 3.1 | 99.9 | -0.7 | 211.445 | 685.42 | 20147 | 127.3 |
| 1/31/2008 | 67.8 | 0.3 | 0.1 | 3.1 | 99.9 | 0.1 | 212.174 | 685.42 | 20147 | 129.2 |
| 2/29/2008 | 67.8 | 0.2 | 0.1 | 3 | 99.9 | -0.6 | 212.687 | 685.42 | 20147 | 129.5 |
| 3/31/2008 | 67.8 | 0.4 | 0.1 | 3 | 99.9 | 0.4 | 213.448 | 685.42 | 31718 | 133.5 |
| 4/30/2008 | 67.8 | 0.2 | 0.1 | 3 | 99.9 | 0.5 | 213.942 | 685.42 | 31718 | 137.3 |
| 5/31/2008 | 67.8 | 0.6 | 0.1 | 3 | 99.9 | 1.4 | 215.208 | 685.42 | 31718 | 141.2 |
| 6/30/2008 | 68.1 | 1 | 0.1 | 3 | 99.9 | 0.8 | 217.463 | 685.42 | 38771 | 145.5 |
| 7/31/2008 | 68.1 | 0.7 | 0.1 | 3.1 | 99.9 | 0.4 | 219.016 | 685.42 | 38771 | 147.5 |
| 8/31/2008 | 68.1 | -0.1 | 0.1 | 3.2 | 99.9 | -1.1 | 218.69 | 685.42 | 38771 | 143 |
| 9/30/2008 | 67.9 | 0.1 | 0.1 | 3.2 | 99.9 | -1 | 218.877 | 685.42 | 29058 | 137.8 |
| 10/31/2008 | 67.9 | -0.9 | 0.1 | 3.1 | 99.9 | -2.6 | 216.995 | 685.42 | 29058 | 129.6 |
| 11/30/2008 | 67.9 | -1.8 | 0.1 | 2.9 | 99.9 | -4 | 213.153 | 685.42 | 29058 | 120 |
| 12/31/2008 | 67.5 | -0.8 | 0.1 | 2.7 | 99.9 | -2.2 | 211.398 | 731.06 | 26175 | 114.5 |
| 1/31/2009 | 67.5 | 0.3 | 0.1 | 2.6 | 99.9 | 1.1 | 211.933 | 731.06 | 26175 | 113 |
| 2/28/2009 | 67.5 | 0.4 | 0.1 | 2.6 | 99.9 | 0.1 | 212.705 | 731.06 | 26175 | 113 |
| 3/31/2009 | 67.3 | -0.1 | 0.1 | 2.5 | 99.9 | -1.7 | 212.495 | 731.06 | 13199 | 113.6 |
| 4/30/2009 | 67.3 | 0.1 | 0.1 | 2.4 | 99.9 | 0.4 | 212.709 | 731.06 | 13199 | 114.8 |
| 5/31/2009 | 67.3 | 0.1 | 0.1 | 2.4 | 99.9 | 0.9 | 213.022 | 731.06 | 13199 | 116.8 |
| 6/30/2009 | 67.4 | 0.8 | 0.1 | 2.1 | 99.9 | 1.1 | 214.79 | 731.06 | 25058 | 120 |
| 7/31/2009 | 67.4 | 0 | 0.1 | 1.8 | 99.9 | -0.2 | 214.726 | 731.06 | 25058 | 119.3 |
| 8/31/2009 | 67.4 | 0.3 | 0.1 | 1.6 | 99.9 | 0.6 | 215.445 | 731.06 | 25058 | 121.1 |
| 9/30/2009 | 67.6 | 0.2 | 0.1 | 1.5 | 99.9 | 0.2 | 215.861 | 731.06 | 25368 | 121.3 |
| 10/31/2009 | 67.6 | 0.3 | 0.1 | 1.4 | 99.9 | 0.1 | 216.509 | 731.06 | 25368 | 122.3 |
| 11/30/2009 | 67.6 | 0.3 | 0.1 | 1.3 | 99.9 | 0.6 | 217.234 | 731.06 | 25368 | 124.1 |
| 12/31/2009 | 67.2 | 0.1 | 0.1 | 1.3 | 99.9 | 0.7 | 217.347 | 748.22 | 34366 | 124.4 |
| 1/31/2010 | 67.2 | 0.1 | 0.9 | 1.1 | 99.9 | 0 | 217.488 | 748.22 | 34366 | 125.9 |
| 2/28/2010 | 67.2 | -0.1 | -0.2 | 0.9 | 99.9 | 0.6 | 217.281 | 748.22 | 34366 | 125.8 |
| 3/31/2010 | 67.1 | 0 | 0.1 | 0.8 | 99.9 | 0.8 | 217.353 | 748.22 | 30539 | 126.3 |
| 4/30/2010 | 67.1 | 0 | 0.3 | 0.6 | 99.9 | 0.8 | 217.403 | 748.22 | 30539 | 127.7 |
| 5/31/2010 | 67.1 | -0.1 | 0.2 | 0.5 | 100.2 | -1 | 217.29 | 748.22 | 30539 | 126.7 |
| 6/30/2010 | 66.9 | 0 | -0.2 | 0.6 | 100.1 | -0.1 | 217.199 | 748.22 | 35994 | 125.2 |
| 7/31/2010 | 66.9 | 0.2 | 0.2 | 0.6 | 100.3 | -0.1 | 217.605 | 748.22 | 35994 | 125.2 |
| 8/31/2010 | 66.9 | 0.1 | 0.2 | 0.5 | 100.3 | 0.6 | 217.923 | 748.22 | 35994 | 125.7 |
| 9/30/2010 | 66.9 | 0.2 | 0.3 | 0.6 | 100.5 | 0.8 | 218.275 | 748.22 | 38404 | 125.7 |
| 10/31/2010 | 66.9 | 0.3 | 0.4 | 0.6 | 100.6 | 0.8 | 219.035 | 748.22 | 38404 | 127.1 |
| 11/30/2010 | 66.9 | 0.3 | 0.3 | 0.6 | 100.7 | 1 | 219.59 | 748.22 | 38404 | 129.2 |
| 12/31/2010 | 66.5 | 0.4 | 0.3 | 0.7 | 100.8 | 0.7 | 220.472 | 728.86 | 40152 | 131 |

| date | USAFCAS | USAPPIMC | UNITEDSTAEMPPER | USAPEFEATSM | UNITEDSTAAPICRURUN | UNITEDSTAAPIGASSTO | UNITEDSTAPERSPE | USAEPY | UNITEDSTAGDPFROMIN | UNITEDSTAECOOPTIND |
|---|---|---|---|---|---|---|---|---|---|---|
| 1/31/2001 | 7312 | 0.5 | 137778 | 0.1 | -0.191 | -1.726 | 0.5 | 1.1 | 263.3 | 53.9 |
| 2/28/2001 | 7312 | 0.2 | 137612 | 0.1 | -0.191 | -1.726 | 0.2 | 0.6 | 263.3 | 53.9 |
| 3/31/2001 | 7312 | 0 | 137783 | 0.1 | -0.191 | -1.726 | -0.1 | 0 | 263.3 | 53.9 |
| 4/30/2001 | 7312 | 0.2 | 137299 | 0.1 | -0.191 | -1.726 | 0.1 | -0.1 | 263.3 | 52.4 |
| 5/31/2001 | 7312 | 0.3 | 137092 | 0.1 | -0.191 | -1.726 | 0.7 | -0.6 | 263.3 | 54.8 |
| 6/30/2001 | 7312 | 0.2 | 136873 | 0.1 | -0.191 | -1.726 | 0.2 | -0.7 | 263.3 | 55.3 |
| 7/31/2001 | 7312 | 0 | 137071 | 0.1 | -0.191 | -1.726 | 0.2 | -1 | 263.3 | 54 |
| 8/31/2001 | 7312 | 0 | 136241 | 0.1 | -0.191 | -1.726 | 0.6 | -1 | 263.3 | 54.4 |
| 9/30/2001 | 7312 | -0.3 | 136846 | 0.1 | -0.191 | -1.726 | -1.6 | -1.4 | 263.3 | 52.1 |
| 10/31/2001 | 7312 | 0.4 | 136392 | 0.1 | -0.191 | -1.726 | 2.9 | -2 | 263.3 | 57.6 |
| 11/30/2001 | 7312 | -0.1 | 136238 | 0.1 | -0.191 | -1.726 | -0.4 | -2.5 | 263.3 | 56.7 |
| 12/31/2001 | 7312 | -0.1 | 136047 | 0.1 | -0.191 | -1.726 | -0.2 | -2.5 | 263.3 | 60.5 |
| 1/31/2002 | 7312 | 0.1 | 135701 | 0.1 | -0.191 | -1.726 | 0.2 | -2.8 | 263.3 | 61.1 |
| 2/28/2002 | 7312 | 0.2 | 136438 | 0.1 | -0.191 | -1.726 | 0.6 | -2.9 | 263.3 | 60.4 |
| 3/31/2002 | 7312 | 0.3 | 136177 | 0.1 | -0.191 | -1.726 | 0.3 | -2.4 | 263.3 | 62.9 |
| 4/30/2002 | 7312 | 0.4 | 136126 | 0.1 | -0.191 | -1.726 | 1 | -1.9 | 263.3 | 60.5 |
| 5/31/2002 | 7312 | 0.1 | 136539 | 0.1 | -0.191 | -1.726 | -0.3 | -1.6 | 263.3 | 60.6 |
| 6/30/2002 | 7312 | 0.1 | 136415 | 0.1 | -0.191 | -1.726 | 0.5 | -1.4 | 263.3 | 59.7 |
| 7/31/2002 | 7312 | 0.2 | 136413 | 0.1 | -0.191 | -1.726 | 0.9 | -0.7 | 263.3 | 54.9 |
| 8/31/2002 | 7312 | 0.2 | 136705 | 0.1 | -0.191 | -1.726 | 0.3 | -0.3 | 263.3 | 55.6 |
| 9/30/2002 | 7312 | 0.2 | 137302 | 0.1 | -0.191 | -1.726 | -0.3 | -0.2 | 263.3 | 54.9 |
| 10/31/2002 | 7312 | 0.2 | 137008 | 0.1 | -0.191 | -1.726 | 0.6 | 0.4 | 263.3 | 53.4 |
| 11/30/2002 | 7312 | 0.1 | 136521 | 0.1 | -0.191 | -1.726 | 0.4 | 1 | 263.3 | 54.4 |
| 12/31/2002 | 7312 | 0.1 | 136426 | 0.1 | -0.191 | -1.726 | 0.8 | 1 | 263.3 | 53.3 |
| 1/31/2003 | 8380 | 0.3 | 137417 | 0.1 | -0.191 | -1.726 | 0.3 | 1.4 | 263.3 | 52.4 |
| 2/28/2003 | 8380 | 0.4 | 137482 | 0.1 | -0.191 | -1.726 | 0.1 | 2.3 | 263.3 | 50 |
| 3/31/2003 | 8380 | 0.3 | 137434 | 0.1 | -0.191 | -1.726 | 0.8 | 2.2 | 263.3 | 48.8 |
| 4/30/2003 | 8380 | -0.2 | 137633 | 0.1 | -0.191 | -1.726 | 0.3 | 1.6 | 263.3 | 56.4 |
| 5/31/2003 | 8380 | -0.1 | 137544 | 0.1 | -0.191 | -1.726 | 0.2 | 1.7 | 263.3 | 56.2 |
| 6/30/2003 | 8380 | 0.1 | 137790 | 0.1 | -0.191 | -1.726 | 0.6 | 1.5 | 263.3 | 56.1 |
| 7/31/2003 | 8363 | 0.3 | 137474 | 0.1 | -0.191 | -1.726 | 0.7 | 1.1 | 263.3 | 56.2 |
| 8/31/2003 | 8380 | 0.3 | 137549 | 0.1 | -0.191 | -1.726 | 1.2 | 0.9 | 263.3 | 54.8 |
| 9/30/2003 | 8380 | 0.3 | 137609 | 0.1 | -0.191 | -1.726 | 0 | 1 | 263.3 | 52.5 |
| 10/31/2003 | 8380 | 0 | 137984 | 0.1 | -0.191 | -1.726 | 0.2 | 1.3 | 263.3 | 53.3 |
| 11/30/2003 | 8380 | 0.1 | 138424 | 0.1 | -0.191 | -1.726 | 0.7 | 1.7 | 263.3 | 55.9 |
| 12/31/2003 | 8847 | 0.2 | 138411 | 0.1 | -0.191 | -1.726 | 0.4 | 2.2 | 263.3 | 58.8 |
| 1/31/2004 | 8847 | 0.4 | 138472 | 0.1 | -0.191 | -1.726 | 0.7 | 2.6 | 263.3 | 60.6 |
| 2/29/2004 | 8700 | 0.2 | 138542 | 0.1 | -0.191 | -1.726 | 0.4 | 2.7 | 263.3 | 56.5 |
| 3/31/2004 | 8847 | 0.2 | 138453 | 0.1 | -0.191 | -1.726 | 0.7 | 3.3 | 263.3 | 54.5 |
| 4/30/2004 | 8794 | 0.2 | 138680 | 0.1 | -0.191 | -1.726 | 0.1 | 4.1 | 263.3 | 52.8 |
| 5/31/2004 | 8847 | 0.3 | 138852 | 0.1 | -0.191 | -1.726 | 0.9 | 4.4 | 263.3 | 54.8 |
| 6/30/2004 | 8847 | 0.3 | 139174 | 0.1 | -0.191 | -1.726 | -0.2 | 3.9 | 263.3 | 52.8 |
| 7/31/2004 | 8779 | 0.1 | 139556 | 0.1 | -0.191 | -1.726 | 1 | 4.5 | 263.3 | 57.3 |
| 8/31/2004 | 8834 | 0.1 | 139573 | 0.1 | -0.191 | -1.726 | 0.4 | 4 | 263.3 | 57.7 |
| 9/30/2004 | 8847 | 0.2 | 139487 | 0.1 | -0.191 | -1.726 | 0.8 | 4 | 263.3 | 54.3 |
| 10/31/2004 | 8847 | 0.4 | 139732 | 0.1 | -0.191 | -1.726 | 0.7 | 4.4 | 263.3 | 57.5 |
| 11/30/2004 | 8847 | 0.4 | 140231 | 0.1 | -0.191 | -1.726 | 0.6 | 4.2 | 263.3 | 55.1 |
| 12/31/2004 | 8847 | 0.1 | 140125 | 0.1 | -0.191 | -1.726 | 0.7 | 4 | 263.3 | 54.5 |
| 1/31/2005 | 10994 | 0.1 | 140245 | 0.1 | -0.191 | -1.726 | -0.1 | 4 | 263.3 | 56.2 |
| 2/28/2005 | 11114 | 0.3 | 140385 | 0.1 | -0.191 | -1.726 | 0.7 | 3.4 | 263.3 | 54.8 |
| 3/31/2005 | 11177 | 0.3 | 140654 | 0.1 | -0.191 | -1.726 | 0.5 | 3.3 | 263.3 | 53 |
| 4/30/2005 | 11357 | 0.3 | 141254 | 0.1 | -0.191 | -1.726 | 0.9 | 3.1 | 263.3 | 47.4 |
| 5/31/2005 | 11417 | 0.1 | 141609 | 0.1 | -0.191 | -1.726 | -0.1 | 2.5 | 263.3 | 47.2 |
| 6/30/2005 | 11456 | 0.1 | 141714 | 0.1 | -0.191 | -1.726 | 0.9 | 3.2 | 262.9 | 50.5 |
| 7/31/2005 | 11513 | 0.4 | 142026 | 0.1 | -0.191 | -1.726 | 1.1 | 2.8 | 262.9 | 48.6 |
| 8/31/2005 | 11630 | 0.4 | 142434 | 0.1 | -0.191 | -1.726 | 0 | 3.1 | 262.9 | 50.9 |
| 9/30/2005 | 11623 | 1 | 142401 | 0.1 | -0.191 | -1.726 | 0.6 | 3.6 | 246.4 | 41.2 |
| 10/31/2005 | 11630 | 0.2 | 142548 | 0.1 | -0.191 | -1.726 | 0.4 | 3.7 | 246.4 | 42 |
| 11/30/2005 | 11630 | -0.2 | 142499 | 0.1 | -0.191 | -1.726 | 0.1 | 2.8 | 246.4 | 48.6 |
| 12/31/2005 | 11630 | 0 | 142752 | 0.1 | -0.191 | -1.726 | 0.4 | 2.8 | 241.4 | 51.1 |
| 1/31/2006 | 12901 | 0.5 | 143150 | 0.1 | -0.191 | -1.726 | 1 | 2.7 | 241.4 | 50.5 |
| 2/28/2006 | 12901 | 0.1 | 143457 | 0.1 | -0.191 | -1.726 | 0.3 | 2.7 | 241.4 | 50.5 |
| 3/31/2006 | 12901 | 0.2 | 143741 | 0.1 | -0.191 | -1.726 | 0.4 | 2.3 | 282.2 | 49.1 |
| 4/30/2006 | 12901 | 0.5 | 143761 | 0.1 | -0.191 | -1.726 | 0.6 | 2.5 | 282.2 | 48.6 |
| 5/31/2006 | 12901 | 0.3 | 144089 | 0.1 | -0.191 | -1.726 | 0.4 | 3.5 | 282.2 | 46.1 |
| 6/30/2006 | 12901 | 0.2 | 144353 | 0.1 | -0.191 | -1.726 | 0.3 | 4.2 | 285.4 | 46.2 |
| 7/31/2006 | 12901 | 0.3 | 144202 | 0.1 | -0.191 | -1.726 | 0.9 | 4.5 | 285.4 | 47.3 |
| 8/31/2006 | 12901 | 0.3 | 144625 | 0.1 | -0.191 | -1.726 | 0 | 5.2 | 285.4 | 45.6 |
| 9/30/2006 | 12901 | -0.3 | 144815 | 0.1 | -0.191 | -1.726 | 0.4 | 3.9 | 292.7 | 50.5 |
| 10/31/2006 | 12901 | -0.2 | 145314 | 0.1 | -0.191 | -1.726 | 0.2 | 2.9 | 292.7 | 52.4 |
| 11/30/2006 | 10554 | 0 | 145534 | 0.1 | -0.191 | -1.726 | 0.1 | 3.9 | 292.7 | 55.7 |
| 12/31/2006 | 10554 | 0.4 | 145970 | 0.1 | -0.191 | -1.726 | 0.9 | 4.5 | 311.1 | 53.5 |
| 1/31/2007 | 15127 | 0.3 | 146028 | 0.1 | -0.191 | -1.726 | 0.5 | 4.1 | 311.1 | 53.7 |
| 2/28/2007 | 15282 | 0.3 | 146057 | 0.1 | -0.191 | -1.726 | 0.3 | 4.9 | 311.1 | 52.7 |
| 3/31/2007 | 15361 | 0.4 | 146320 | 0.1 | -0.191 | -1.726 | 0.5 | 5.4 | 316.7 | 50.8 |
| 4/30/2007 | 15362 | 0.2 | 145586 | 0.1 | -0.191 | -1.726 | 0.3 | 5.1 | 316.7 | 45.5 |
| 5/31/2007 | 15375 | 0.3 | 145903 | 0.1 | -0.191 | -1.726 | 0.4 | 4.6 | 316.7 | 48 |
| 6/30/2007 | 15398 | 0.2 | 146063 | 0.1 | -0.191 | -1.726 | 0.2 | 4.3 | 325.5 | 49.1 |
| 7/31/2007 | 15399 | 0.2 | 145905 | 0.1 | -0.191 | -1.726 | 0.5 | 4 | 325.5 | 48.2 |
| 8/31/2007 | 15423 | 0.1 | 145682 | 0.1 | -0.191 | -1.726 | 0.5 | 3.7 | 325.5 | 49.5 |
| 9/30/2007 | 15425 | 0.4 | 146244 | 0.1 | -0.191 | -1.726 | 0.5 | 4.5 | 318.8 | 48.2 |
| 10/31/2007 | 15455 | 0.3 | 145946 | 0.1 | -0.191 | -1.726 | 0.4 | 5.6 | 318.8 | 47.3 |
| 11/30/2007 | 15457 | 0.5 | 146595 | 0.1 | -0.191 | -1.726 | 0.7 | 6.2 | 318.8 | 43.8 |

| date | | | | | | | | | |
|---|---|---|---|---|---|---|---|---|---|
| 12/31/2007 | 15462 | 0.2 | 146273 | 0.1 | -0.191 | -1.726 | 0.2 | 6 | 302.6 | 44.4 |
| 1/31/2008 | 18480 | 0.3 | 146378 | 0.1 | -0.191 | -1.726 | 0.2 | 6.8 | 302.6 | 43.2 |
| 2/29/2008 | 18457 | 0.2 | 146156 | 0.1 | -0.191 | -1.726 | -0.1 | 6.9 | 302.6 | 44.5 |
| 3/31/2008 | 18458 | 0.3 | 146086 | 0.1 | -0.191 | -1.726 | 0.5 | 7.9 | 285.8 | 42.5 |
| 4/30/2008 | 18476 | 0.2 | 146132 | 0.1 | -0.191 | -1.726 | 0.3 | 8 | 285.8 | 39.2 |
| 5/31/2008 | 18479 | 0.4 | 145908 | 0.1 | -0.191 | -1.726 | 0.6 | 8.1 | 285.8 | 40.3 |
| 6/30/2008 | 18486 | 0.7 | 145737 | 0.1 | -0.191 | -1.726 | 0.5 | 8.7 | 275.2 | 37.4 |
| 7/31/2008 | 18501 | 0.5 | 145532 | 0.1 | -0.191 | -1.726 | 0 | 10.2 | 275.2 | 37.4 |
| 8/31/2008 | 18509 | -0.1 | 145203 | 0.1 | -0.191 | -1.726 | -0.1 | 8.3 | 275.2 | 42.8 |
| 9/30/2008 | 18516 | 0.1 | 145076 | 0.1 | -0.191 | -1.726 | -0.6 | 7 | 291.2 | 45.8 |
| 10/31/2008 | 18335 | -0.7 | 144802 | 0.1 | -0.191 | -1.726 | -0.8 | 4 | 291.2 | 41.1 |
| 11/30/2008 | 17170 | -1.2 | 144100 | 0.1 | -0.191 | -1.726 | -1.5 | -0.3 | 291.2 | 50.8 |
| 12/31/2008 | 21076 | -0.6 | 143369 | 0.1 | -0.191 | -1.726 | -0.8 | -2.9 | 371.2 | 45 |
| 1/31/2009 | 18977 | 0.9 | 142152 | 0.1 | -0.191 | -1.726 | 0.5 | -3.4 | 371.2 | 45.4 |
| 2/28/2009 | 19832 | -0.7 | 141640 | 0.1 | -0.191 | -1.726 | -0.2 | -4.5 | 371.2 | 44.6 |
| 3/31/2009 | 21148 | -0.1 | 140707 | 0.1 | -0.191 | -1.726 | -0.5 | -6.7 | 387.4 | 45.3 |
| 4/30/2009 | 21190 | 0.1 | 140656 | 0.1 | -0.191 | -1.726 | 0 | -6.7 | 387.4 | 49.1 |
| 5/31/2009 | 19174 | 0.1 | 140248 | 0.1 | -0.191 | -1.726 | 0.2 | -6.6 | 387.4 | 48.6 |
| 6/30/2009 | 21256 | 0.6 | 140009 | 0.1 | -0.191 | -1.726 | 0.6 | -6.6 | 353.9 | 50.8 |
| 7/31/2009 | 21297 | 0 | 139901 | 0.1 | -0.191 | -1.726 | 0.3 | -8.3 | 353.9 | 46.3 |
| 8/31/2009 | 21331 | 0.3 | 139492 | 0.1 | -0.191 | -1.726 | 1.2 | -6.2 | 353.9 | 50.3 |
| 9/30/2009 | 21373 | 0.2 | 138818 | 0.1 | -0.191 | -1.726 | -0.8 | -5.6 | 338.6 | 52.5 |
| 10/31/2009 | 21406 | 0.4 | 138432 | 0.1 | -0.191 | -1.726 | 0.5 | -3.6 | 338.6 | 48.7 |
| 11/30/2009 | 21440 | 0.2 | 138659 | 0.1 | -0.191 | -1.726 | 0.1 | 0.4 | 338.6 | 47.9 |
| 12/31/2009 | 21482 | 0.1 | 138013 | 0.1 | -0.191 | -1.726 | 0.7 | 3.4 | 319.1 | 46.8 |
| 1/31/2010 | 25294 | 0.6 | 138438 | 0.1 | -0.191 | -1.726 | 0 | 3.5 | 319.1 | 48.8 |
| 2/28/2010 | 25275 | 0 | 138581 | 0.1 | -0.191 | -1.726 | 0.4 | 3.4 | 319.1 | 46.8 |
| 3/31/2010 | 24845 | 0.1 | 138751 | 0.1 | -0.191 | -1.726 | 0.6 | 4.9 | 301.2 | 45.4 |
| 4/30/2010 | 25589 | 0.1 | 139297 | 0.1 | -0.191 | -1.726 | 0.3 | 5.5 | 301.2 | 48.4 |
| 5/31/2010 | 25668 | 0 | 139241 | 0.1 | -0.191 | -1.726 | 0.3 | 5.6 | 301.2 | 48.7 |
| 6/30/2010 | 25798 | 0 | 139141 | 0.1 | -0.191 | -1.726 | 0.2 | 3.7 | 310 | 46.2 |
| 7/31/2010 | 25822 | 0.1 | 139179 | 0.1 | -0.191 | -1.726 | 0.4 | 3.9 | 310 | 44.7 |
| 8/31/2010 | 25845 | 0.1 | 139438 | 0.1 | -0.191 | -1.726 | 0.4 | 4.1 | 310 | 43.6 |
| 9/30/2010 | 25875 | 0.1 | 139396 | 0.1 | -0.191 | -1.726 | 0.2 | 4.9 | 310.5 | 45.3 |
| 10/31/2010 | 25898 | 0.3 | 139119 | 0.1 | -0.191 | -1.726 | 0.6 | 5.8 | 310.5 | 46.4 |
| 11/30/2010 | 25921 | 0.2 | 139044 | 0.1 | -0.191 | -1.726 | 0.5 | 6.5 | 310.5 | 46.7 |
| 12/31/2010 | 25951 | 0.2 | 139301 | 0.1 | -0.191 | -1.726 | 0.4 | 6.5 | 318.1 | 45.8 |

| date | UNITEDSTAHOMOWNRAT | UNITEDSTAINFRATMOM | USAPPIM | UNITEDSTACPIMED | UNITEDSTACORPROPRI | UNITEDSTARETSALEXAUT | USACSA | USAPREC | UNITEDSTAFORDIRINV | UNITEDSTAIMPPRI |
|---|---|---|---|---|---|---|---|---|---|---|
| 1/31/2011 | 66.5 | 0.3 | 0.6 | 0.9 | 101.2 | 0.6 | 221.187 | 728.86 | 40152 | 133 |
| 2/28/2011 | 66.5 | 0.3 | 0.6 | 1.1 | 101.5 | 0.7 | 221.898 | 728.86 | 40152 | 135.3 |
| 3/31/2011 | 66.4 | 0.5 | 0.7 | 1.3 | 102 | 1 | 223.046 | 728.86 | 39368 | 139.3 |
| 4/30/2011 | 66.4 | 0.5 | 0.5 | 1.4 | 102.3 | 0.9 | 224.093 | 728.86 | 39368 | 142.9 |
| 5/31/2011 | 66.4 | 0.3 | 0.3 | 1.6 | 102.5 | 0.2 | 224.806 | 728.86 | 39368 | 143.1 |
| 6/30/2011 | 65.9 | 0 | 0.1 | 1.6 | 102.7 | 0.7 | 224.806 | 728.86 | 46165 | 142.2 |
| 7/31/2011 | 65.9 | 0.3 | 0.2 | 1.8 | 102.9 | 0 | 225.395 | 728.86 | 46165 | 142.4 |
| 8/31/2011 | 65.9 | 0.3 | 0.2 | 2 | 103.3 | 0.5 | 226.106 | 728.86 | 46165 | 141.9 |
| 9/30/2011 | 66.3 | 0.2 | 0.4 | 2.1 | 103.5 | 0.3 | 226.597 | 728.86 | 43580 | 141.7 |
| 10/31/2011 | 66.3 | 0.1 | -0.4 | 2.2 | 103.2 | 0.4 | 226.75 | 728.86 | 43580 | 141.2 |
| 11/30/2011 | 66.3 | 0.2 | 0.3 | 2.2 | 103.4 | 0.3 | 227.169 | 728.86 | 43580 | 142.2 |
| 12/31/2011 | 66 | 0 | -0.1 | 2.3 | 103.4 | -0.3 | 227.223 | 730.03 | 36855 | 142.2 |
| 1/31/2012 | 66 | 0.3 | 0.4 | 2.4 | 103.8 | 1.2 | 227.842 | 730.03 | 36855 | 142.2 |
| 2/29/2012 | 66 | 0.2 | 0.3 | 2.3 | 104.1 | 1.2 | 228.329 | 730.03 | 36855 | 142.2 |
| 3/31/2012 | 65.4 | 0.2 | 0.2 | 2.4 | 104.3 | 0.4 | 228.807 | 730.03 | 39347 | 144.2 |
| 4/30/2012 | 65.4 | 0.2 | 0.3 | 2.4 | 104.6 | -0.6 | 229.187 | 730.03 | 39347 | 144.1 |
| 5/31/2012 | 65.4 | -0.2 | -0.1 | 2.3 | 104.8 | -0.5 | 228.713 | 730.03 | 39347 | 142 |
| 6/30/2012 | 65.5 | -0.1 | -0.3 | 2.3 | 104.7 | -0.7 | 228.524 | 730.03 | 39354 | 138.7 |
| 7/31/2012 | 65.5 | 0 | -0.1 | 2.3 | 104.6 | 0.3 | 228.59 | 730.03 | 39354 | 137.7 |
| 8/31/2012 | 65.5 | 0.6 | 0.3 | 2.2 | 104.5 | 1.2 | 229.918 | 730.03 | 39354 | 139.4 |
| 9/30/2012 | 65.5 | 0.5 | 0.7 | 2.2 | 104.8 | 0.6 | 231.015 | 730.03 | 42695 | 140.8 |
| 10/31/2012 | 65.5 | 0.3 | 0.1 | 2.2 | 104.9 | 0.1 | 231.638 | 730.03 | 42695 | 141.2 |
| 11/30/2012 | 65.5 | -0.2 | 0.1 | 2.2 | 105.2 | 0.3 | 231.249 | 730.03 | 42695 | 140.2 |
| 12/31/2012 | 65.4 | 0 | 0 | 2.1 | 105.5 | 0.1 | 231.221 | 680.94 | 38932 | 139.4 |
| 1/31/2013 | 65.4 | 0.2 | 0.3 | 2.1 | 105.6 | 0.8 | 231.679 | 680.94 | 38932 | 140.1 |
| 2/28/2013 | 65.4 | 0.5 | 0.2 | 2.2 | 105.6 | 1.1 | 232.937 | 680.94 | 38932 | 141.3 |
| 3/31/2013 | 65 | -0.3 | 0 | 2.1 | 105.8 | -0.7 | 232.282 | 680.94 | 40568 | 141.2 |
| 4/30/2013 | 65 | -0.2 | -0.2 | 2.1 | 106 | -0.7 | 231.797 | 680.94 | 40568 | 140.2 |
| 5/31/2013 | 65 | 0 | -0.1 | 2.1 | 105.7 | 0.3 | 231.893 | 680.94 | 40568 | 139.4 |
| 6/30/2013 | 65 | 0.2 | 0.4 | 2.1 | 106.1 | -0.1 | 232.445 | 680.94 | 42779 | 138.8 |
| 7/31/2013 | 65 | 0.2 | 0.2 | 2.1 | 106.4 | 0.7 | 232.9 | 680.94 | 42779 | 138.9 |
| 8/31/2013 | 65 | 0.2 | 0.1 | 2.1 | 106.4 | -0.1 | 233.456 | 680.94 | 42779 | 139.4 |
| 9/30/2013 | 65.3 | 0 | 0 | 2.1 | 106.5 | 0.4 | 233.544 | 680.94 | 41796 | 139.8 |
| 10/31/2013 | 65.3 | 0.1 | 0.2 | 2 | 106.7 | 0.1 | 233.669 | 680.94 | 41796 | 138.9 |
| 11/30/2013 | 65.3 | 0.2 | 0.2 | 2 | 106.9 | 0.3 | 234.1 | 680.94 | 41796 | 137.7 |
| 12/31/2013 | 65.2 | 0.3 | 0.1 | 2.1 | 106.9 | 0.8 | 234.719 | 768.38 | 43092 | 137.8 |
| 1/31/2014 | 65.2 | 0.2 | 0.3 | 2.1 | 107.1 | -0.4 | 235.288 | 768.38 | 43092 | 138.3 |
| 2/28/2014 | 65.2 | 0.1 | 0.2 | 2.1 | 107.3 | 1 | 235.547 | 768.38 | 43092 | 139.8 |
| 3/31/2014 | 64.8 | 0.2 | 0.4 | 2.2 | 107.6 | 0.4 | 236.028 | 768.38 | 38971 | 140.5 |
| 4/30/2014 | 64.8 | 0.2 | 0.1 | 2.2 | 107.6 | 0.9 | 236.468 | 768.38 | 38971 | 139.7 |
| 5/31/2014 | 64.8 | 0.2 | 0.2 | 2.3 | 107.9 | 0.2 | 236.918 | 768.38 | 38971 | 140.1 |
| 6/30/2014 | 64.7 | 0.1 | -0.1 | 2.2 | 107.9 | 0.3 | 237.231 | 768.38 | 45662 | 140.5 |
| 7/31/2014 | 64.7 | 0.1 | 0.4 | 2.3 | 108.4 | 0.1 | 237.498 | 768.38 | 45662 | 140.1 |

| Date | | | | | | | | | |
|---|---|---|---|---|---|---|---|---|---|
| 8/31/2014 | 64.7 | 0 | 0 | 2.2 | 108.4 | 0.7 | 237.46 | 768.38 | 45662 | 139 |
| 9/30/2014 | 64.4 | 0 | -0.2 | 2.2 | 108.3 | 0 | 237.477 | 768.38 | 45196 | 137.9 |
| 10/31/2014 | 64.4 | 0 | 0.2 | 2.3 | 108.7 | 0.3 | 237.43 | 768.38 | 45196 | 136 |
| 11/30/2014 | 64.4 | -0.2 | -0.2 | 2.3 | 108.7 | 0 | 236.983 | 768.38 | 45196 | 133.5 |
| 12/31/2014 | 64 | -0.3 | -0.3 | 2.2 | 108.9 | -0.7 | 236.252 | 742.1 | 47465 | 130.1 |
| 1/31/2015 | 64 | -0.6 | -0.6 | 2.2 | 108.9 | -1 | 234.747 | 742.1 | 47465 | 126 |
| 2/28/2015 | 64 | 0.3 | -0.5 | 2.2 | 108.5 | 0.2 | 235.342 | 742.1 | 47465 | 125.5 |
| 3/31/2015 | 63.7 | 0.3 | 0.2 | 2.1 | 108.6 | 0.8 | 235.976 | 742.1 | 34555 | 125.3 |
| 4/30/2015 | 63.7 | 0.1 | -0.2 | 2.2 | 108.6 | 0.1 | 236.222 | 742.1 | 34555 | 125.1 |
| 5/31/2015 | 63.7 | 0.3 | 0.4 | 2.2 | 108.7 | 0.7 | 237.001 | 742.1 | 34555 | 126.5 |
| 6/30/2015 | 63.4 | 0.3 | 0.3 | 2.3 | 109 | 0.1 | 237.657 | 742.1 | 42723 | 126.6 |
| 7/31/2015 | 63.4 | 0.2 | 0.2 | 2.3 | 109.2 | 0.9 | 238.034 | 742.1 | 42723 | 125.4 |
| 8/31/2015 | 63.4 | 0 | -0.2 | 2.3 | 109.2 | -0.1 | 238.033 | 742.1 | 42723 | 123.2 |
| 9/30/2015 | 63.7 | -0.2 | -0.4 | 2.4 | 109.1 | -0.7 | 237.498 | 742.1 | 43163 | 121.9 |
| 10/31/2015 | 63.7 | 0.1 | -0.3 | 2.4 | 108.9 | -0.1 | 237.733 | 742.1 | 43163 | 121.5 |
| 11/30/2015 | 63.7 | 0.1 | 0.1 | 2.4 | 109 | 0.2 | 238.017 | 742.1 | 43163 | 120.8 |
| 12/31/2015 | 63.8 | -0.1 | -0.1 | 2.3 | 109.2 | 0.6 | 237.761 | 829.56 | 29484 | 119.3 |
| 1/31/2016 | 63.8 | 0 | 0.4 | 2.4 | 109.8 | -0.8 | 237.652 | 829.56 | 29484 | 117.8 |
| 2/29/2016 | 63.8 | -0.1 | -0.3 | 2.4 | 109.8 | 0.6 | 237.336 | 829.56 | 29484 | 117.2 |
| 3/31/2016 | 63.5 | 0.3 | 0 | 2.4 | 109.7 | 0.1 | 238.08 | 829.56 | 31187 | 117.7 |
| 4/30/2016 | 63.5 | 0.4 | 0.2 | 2.5 | 109.9 | 0.4 | 238.992 | 829.56 | 31187 | 118.5 |
| 5/31/2016 | 63.5 | 0.2 | 0.2 | 2.5 | 109.9 | 0.4 | 239.557 | 829.56 | 31187 | 119.9 |
| 6/30/2016 | 62.9 | 0.3 | 0.5 | 2.5 | 110.3 | 1.2 | 240.222 | 829.56 | 38558 | 120.7 |
| 7/31/2016 | 62.9 | -0.1 | -0.1 | 2.5 | 110.2 | -0.6 | 240.101 | 829.56 | 38558 | 120.8 |
| 8/31/2016 | 62.9 | 0.2 | -0.2 | 2.6 | 110.2 | 0 | 240.545 | 829.56 | 38558 | 120.5 |
| 9/30/2016 | 63.5 | 0.3 | 0.3 | 2.6 | 110.4 | 0.6 | 241.176 | 829.56 | 38943 | 120.6 |
| 10/31/2016 | 63.5 | 0.2 | 0.3 | 2.5 | 110.5 | 0.4 | 241.741 | 829.56 | 38943 | 121.2 |
| 11/30/2016 | 63.5 | 0.1 | 0.2 | 2.6 | 110.8 | 0.1 | 242.026 | 829.56 | 38943 | 121.1 |
| 12/31/2016 | 63.7 | 0.3 | 0.3 | 2.6 | 110.9 | 0.8 | 242.637 | 749.64 | 32914 | 121.6 |
| 1/31/2017 | 63.7 | 0.4 | 0.4 | 2.6 | 111.3 | 1.4 | 243.618 | 749.64 | 32914 | 122.3 |
| 2/28/2017 | 63.7 | 0.2 | 0 | 2.6 | 111.2 | 0.1 | 244.006 | 749.64 | 32914 | 122.7 |
| 3/31/2017 | 63.6 | 0 | 0.2 | 2.6 | 111.4 | 0.3 | 243.892 | 749.64 | 36272 | 122.5 |
| 4/30/2017 | 63.6 | 0.1 | 0.4 | 2.4 | 111.9 | 0.3 | 244.193 | 749.64 | 36272 | 122.8 |
| 5/31/2017 | 63.6 | -0.1 | -0.1 | 2.3 | 112.2 | -0.7 | 244.004 | 749.64 | 36272 | 122.7 |
| 6/30/2017 | 63.7 | 0.1 | 0.1 | 2.3 | 112.3 | 0.3 | 244.163 | 749.64 | 43589 | 122.4 |
| 7/31/2017 | 63.7 | 0 | 0.1 | 2.3 | 112.4 | 0 | 244.243 | 749.64 | 43589 | 122.2 |
| 8/31/2017 | 63.7 | 0.4 | 0.4 | 2.3 | 112.6 | 0.8 | 245.183 | 749.64 | 43589 | 122.9 |
| 9/30/2017 | 63.9 | 0.5 | 0.4 | 2.3 | 112.7 | 1.5 | 246.435 | 749.64 | 45681 | 123.9 |
| 10/31/2017 | 63.9 | 0.1 | 0.4 | 2.3 | 113.2 | 0 | 246.626 | 749.64 | 45681 | 124.1 |
| 11/30/2017 | 63.9 | 0.3 | 0.4 | 2.3 | 113.4 | 1.6 | 247.284 | 749.64 | 45681 | 125.3 |
| 12/31/2017 | 64.2 | 0.2 | 0.1 | 2.4 | 113.4 | 0.7 | 247.805 | 779.7 | 51091 | 125.5 |
| 1/31/2018 | 64.2 | 0.4 | 0.3 | 2.4 | 113.8 | -0.4 | 248.859 | 779.7 | 51091 | 126.5 |
| 2/28/2018 | 64.2 | 0.3 | 0.3 | 2.4 | 114 | 0.8 | 249.529 | 779.7 | 51091 | 126.8 |
| 3/31/2018 | 64.2 | 0 | 0.2 | 2.4 | 114.4 | -0.1 | 249.577 | 779.7 | 43048 | 126.5 |
| 4/30/2018 | 64.2 | 0.3 | 0.2 | 2.5 | 114.5 | 0.4 | 250.227 | 779.7 | 43048 | 127.1 |
| 5/31/2018 | 64.2 | 0.2 | 0.3 | 2.6 | 114.8 | 1.4 | 250.792 | 779.7 | 43048 | 128.2 |
| 6/30/2018 | 64.3 | 0.1 | 0.3 | 2.7 | 115.3 | -0.3 | 251.018 | 779.7 | 53911 | 128.2 |
| 7/31/2018 | 64.3 | 0.1 | 0.1 | 2.7 | 115.5 | 0.6 | 251.214 | 779.7 | 53911 | 128.1 |
| 8/31/2018 | 64.3 | 0.2 | 0.1 | 2.7 | 115.5 | 0.1 | 251.663 | 779.7 | 53911 | 127.6 |
| 9/30/2018 | 64.4 | 0.2 | 0.2 | 2.7 | 115.6 | -0.3 | 252.182 | 779.7 | 54508 | 127.7 |
| 10/31/2018 | 64.4 | 0.2 | 0.7 | 2.6 | 116.4 | 1.2 | 252.772 | 779.7 | 54508 | 128.3 |
| 11/30/2018 | 64.4 | -0.1 | -0.1 | 2.8 | 116.6 | 0.2 | 252.594 | 779.7 | 54508 | 126.2 |
| 12/31/2018 | 64.8 | 0.1 | 0 | 2.8 | 116.7 | -2.8 | 252.767 | 822.81 | 52221 | 124.4 |
| 1/31/2019 | 64.8 | -0.1 | -0.3 | 2.7 | 116.7 | 1.6 | 252.561 | 822.81 | 52221 | 124.6 |
| 2/28/2019 | 64.8 | 0.3 | 0.2 | 2.8 | 116.9 | -0.1 | 253.319 | 822.81 | 52221 | 125.9 |
| 3/31/2019 | 64.2 | 0.4 | 0.3 | 2.8 | 117 | 1.3 | 254.277 | 822.81 | 45893 | 126.6 |
| 4/30/2019 | 64.2 | 0.4 | 0.5 | 2.8 | 117.5 | 0.6 | 255.233 | 822.81 | 45893 | 126.8 |
| 5/31/2019 | 64.2 | 0 | 0.1 | 2.8 | 117.7 | 0.4 | 255.296 | 822.81 | 45893 | 127 |
| 6/30/2019 | 64.1 | 0 | -0.1 | 2.9 | 117.7 | 0.3 | 255.213 | 822.81 | 50294 | 125.6 |
| 7/31/2019 | 64.1 | 0.2 | 0.3 | 2.9 | 118 | 0.9 | 255.802 | 822.81 | 50294 | 125.6 |
| 8/31/2019 | 64.1 | 0.1 | 0.1 | 2.9 | 118.2 | 0.2 | 256.036 | 822.81 | 50294 | 124.9 |
| 9/30/2019 | 64.8 | 0.2 | -0.3 | 3 | 117.9 | -0.2 | 256.43 | 822.81 | 48295 | 125 |
| 10/31/2019 | 64.8 | 0.3 | 0.4 | 3 | 118.2 | 0.4 | 257.155 | 822.81 | 48295 | 124.5 |
| 11/30/2019 | 64.8 | 0.3 | -0.2 | 3 | 118 | 0.1 | 257.879 | 822.81 | 48295 | 124.7 |
| 12/31/2019 | 65.1 | 0.3 | 0.3 | 2.9 | 118.2 | 0.9 | 258.63 | 837 | 52125 | 125 |
| 1/31/2020 | 65.1 | 0.1 | 0.3 | 2.8 | 118.6 | 0.5 | 258.906 | 837 | 52125 | 125.2 |
| 2/29/2020 | 65.1 | 0.1 | -0.6 | 2.8 | 118.2 | -0.2 | 259.246 | 837 | 52125 | 124.3 |
| 3/31/2020 | 65.3 | -0.4 | -0.5 | 2.8 | 118.3 | -4.9 | 258.15 | 837 | 30441 | 121.3 |
| 4/30/2020 | 65.3 | -0.8 | -1.2 | 2.7 | 117.9 | -14.7 | 256.126 | 837 | 30441 | 118.2 |
| 5/31/2020 | 65.3 | -0.1 | 0.5 | 2.7 | 117.9 | 12.8 | 255.848 | 837 | 30441 | 119 |
| 6/30/2020 | 67.9 | 0.5 | 0.3 | 2.5 | 118.1 | 8.7 | 257.004 | 837 | 24720 | 120.6 |
| 7/31/2020 | 67.9 | 0.5 | 0.6 | 2.6 | 118.7 | 2.1 | 258.408 | 837 | 24720 | 122.1 |
| 8/31/2020 | 67.9 | 0.4 | 0.2 | 2.5 | 118.9 | 0.9 | 259.366 | 837 | 24720 | 123.2 |
| 9/30/2020 | 67.4 | 0.2 | 0.3 | 2.4 | 119.1 | 1.6 | 259.951 | 837 | 44657 | 123.4 |
| 10/31/2020 | 67.4 | 0.1 | 0.6 | 2.4 | 119.7 | -0.3 | 260.249 | 837 | 44657 | 123.3 |
| 11/30/2020 | 67.4 | 0.2 | 0.1 | 2.2 | 119.7 | -0.9 | 260.895 | 837 | 44657 | 123.4 |
| 12/31/2020 | 65.8 | 0.4 | 0.4 | 2.2 | 119.8 | 0.6 | 262.005 | 741.14 | 53704 | 124.6 |

| date | USAFCAS | USAPPIMC | UNITEDSTAEMPPER | USAPEFEATSM | UNITEDSTAAPICRURUN | UNITEDSTAAPIGASSTO | UNITEDSTAPERSPE | USAEPY | UNITEDSTAGDPFROMIN | UNITEDSTAECOOPTIND |
|---|---|---|---|---|---|---|---|---|---|---|
| 1/31/2011 | 26532 | 0.3 | 139250 | 0.1 | -0.191 | -1.726 | 0.4 | 7 | 318.1 | 51.9 |
| 2/28/2011 | 26517 | 0.3 | 139394 | 0.1 | -0.191 | -1.726 | 0.3 | 8.7 | 318.1 | 50.9 |
| 3/31/2011 | 26291 | 0.4 | 139639 | 0.1 | -0.191 | -1.726 | 0.8 | 9.5 | 309.2 | 43 |
| 4/30/2011 | 26285 | 0.4 | 139586 | 0.1 | -0.191 | -1.726 | 0.3 | 9.2 | 309.2 | 40.8 |
| 5/31/2011 | 26329 | 0.3 | 139624 | 0.1 | -0.191 | -1.726 | 0.2 | 9.1 | 309.2 | 42.8 |
| 6/30/2011 | 26487 | 0 | 139384 | 0.1 | -0.191 | -1.726 | 0.2 | 10.1 | 312.4 | 44.6 |
| 7/31/2011 | 25836 | 0.2 | 139524 | 0.1 | -0.191 | -1.726 | 0.3 | 9.8 | 312.4 | 41.4 |
| 8/31/2011 | 25945 | 0.2 | 139942 | 0.1 | -0.191 | -1.726 | 0.2 | 9.4 | 312.4 | 35.8 |
| 9/30/2011 | 26004 | 0.1 | 140183 | 0.1 | -0.191 | -1.726 | 0.3 | 9.4 | 323.8 | 39.9 |
| 10/31/2011 | 26028 | 0 | 140368 | 0.1 | -0.191 | -1.726 | 0.2 | 6.3 | 323.8 | 40.3 |
| 11/30/2011 | 26981 | 0.2 | 140826 | 0.1 | -0.191 | -1.726 | 0 | 4.8 | 323.8 | 40.6 |
| 12/31/2011 | 26900 | 0.1 | 140902 | 0.1 | -0.191 | -1.726 | 0.1 | 3.6 | 343.7 | 42.8 |
| 1/31/2012 | 26901 | 0.3 | 141584 | 0.1 | -0.191 | -1.726 | 0.7 | 2.6 | 343.7 | 47.5 |
| 2/29/2012 | 27260 | 0.2 | 141858 | 0.1 | -0.191 | -1.726 | 0.8 | 1.8 | 343.7 | 49.4 |
| 3/31/2012 | 27217 | 0.2 | 142036 | 0.1 | -0.191 | -1.726 | 0.1 | 1.1 | 349.2 | 47.5 |
| 4/30/2012 | 27231 | 0.1 | 141899 | 0.1 | -0.191 | -1.726 | 0.2 | 0.7 | 349.2 | 49.3 |
| 5/31/2012 | 27329 | -0.1 | 142206 | 0.1 | -0.191 | -1.726 | -0.1 | -0.2 | 349.2 | 48.5 |
| 6/30/2012 | 27334 | -0.1 | 142391 | 0.1 | -0.191 | -1.726 | -0.2 | -2.1 | 364.6 | 46.7 |
| 7/31/2012 | 27339 | 0 | 142292 | 0.1 | -0.191 | -1.726 | 0.2 | -1.3 | 364.6 | 47 |
| 8/31/2012 | 27334 | 0.3 | 142291 | 0.1 | -0.191 | -1.726 | 0.4 | -0.9 | 364.6 | 45.6 |
| 9/30/2012 | 27359 | 0.3 | 143044 | 0.1 | -0.191 | -1.726 | 0.4 | -0.6 | 361.2 | 51.8 |
| 10/31/2012 | 27380 | 0.3 | 143431 | 0.1 | -0.191 | -1.726 | 0.3 | 1.5 | 361.2 | 54 |
| 11/30/2012 | 27590 | -0.1 | 143333 | 0.1 | -0.191 | -1.726 | 0.4 | 0.8 | 361.2 | 48.6 |
| 12/31/2012 | 27364 | 0 | 143330 | 0.1 | -0.191 | -1.726 | 0 | 1.1 | 367.1 | 45.1 |
| 1/31/2013 | 27394 | 0.2 | 143292 | 0.1 | -0.191 | -1.726 | 0.6 | 1.2 | 367.1 | 46.5 |
| 2/28/2013 | 27494 | 0.4 | 143362 | 0.1 | -0.191 | -1.726 | 0.3 | 1.5 | 367.1 | 47.3 |
| 3/31/2013 | 27564 | -0.1 | 143316 | 0.1 | -0.191 | -1.726 | 0 | 0.2 | 364.7 | 42.2 |
| 4/30/2013 | 27564 | -0.1 | 143635 | 0.1 | -0.191 | -1.726 | -0.1 | -0.8 | 364.7 | 46.2 |
| 5/31/2013 | 27602 | 0.1 | 143882 | 0.1 | -0.191 | -1.726 | 0.4 | -0.8 | 364.7 | 45.1 |
| 6/30/2013 | 27488 | 0.2 | 143999 | 0.1 | -0.191 | -1.726 | 0.2 | 0.8 | 368.8 | 49 |
| 7/31/2013 | 27512 | 0.1 | 144264 | 0.1 | -0.191 | -1.726 | 0.2 | 0.3 | 368.8 | 47.1 |
| 8/31/2013 | 27477 | 0.1 | 144326 | 0.1 | -0.191 | -1.726 | 0.3 | -1.1 | 368.8 | 45.1 |
| 9/30/2013 | 27441 | 0 | 144418 | 0.1 | -0.191 | -1.726 | 0.3 | -1.6 | 374.4 | 46 |
| 10/31/2013 | 27427 | 0.1 | 143537 | 0.2 | -0.191 | -1.726 | 0.5 | -2.2 | 374.4 | 38.4 |
| 11/30/2013 | 27482 | 0.1 | 144479 | 0.1 | -0.191 | -1.726 | 0.6 | -1.5 | 374.4 | 41.4 |
| 12/31/2013 | 27509 | 0.2 | 144778 | 0.3 | -0.191 | -1.726 | 0.3 | -1 | 388.3 | 43.1 |
| 1/31/2014 | 27516 | 0.2 | 145150 | 0.2 | -0.191 | -1.726 | -0.1 | -1 | 388.3 | 45.2 |
| 2/28/2014 | 27939 | 0 | 145134 | 0 | -0.191 | -1.726 | 0.5 | -1 | 388.3 | 44.9 |
| 3/31/2014 | 28050 | 0.2 | 145648 | 0.2 | -0.191 | -1.726 | 0.7 | 0.4 | 378.5 | 45.1 |
| 4/30/2014 | 28064 | 0.2 | 145667 | 0.2 | -0.191 | -1.726 | 0.3 | -0.1 | 378.5 | 48 |
| 5/31/2014 | 28171 | 0.2 | 145825 | 0 | -0.191 | -1.726 | 0.4 | 0.6 | 378.5 | 45.8 |
| 6/30/2014 | 28161 | 0.1 | 146247 | 0.1 | -0.191 | -1.726 | 0.5 | 0.2 | 399.9 | 47.7 |
| 7/31/2014 | 28163 | 0.1 | 146399 | 0.1 | -0.191 | -1.726 | 0.4 | 0.4 | 399.9 | 45.6 |
| 8/31/2014 | 28160 | 0 | 146530 | 0.3 | -0.191 | -1.726 | 0.7 | 0.4 | 399.9 | 44.5 |
| 9/30/2014 | 28185 | 0 | 146778 | 0 | -0.191 | -1.726 | 0.1 | -0.4 | 424.8 | 45.2 |
| 10/31/2014 | 28228 | 0 | 147427 | 0.1 | -0.191 | -1.726 | 0.6 | -0.7 | 424.8 | 45.2 |
| 11/30/2014 | 28292 | -0.1 | 147404 | 0 | -0.191 | -1.726 | 0.2 | -1.7 | 424.8 | 46.4 |
| 12/31/2014 | 28572 | -0.2 | 147615 | 0.1 | -0.191 | -1.726 | 0.1 | -3 | 454.9 | 48.4 |
| 1/31/2015 | 28573 | -0.5 | 148145 | -0.1 | -0.191 | -1.726 | -0.2 | -5 | 454.9 | 51.5 |
| 2/28/2015 | 28779 | 0.2 | 148045 | -0.1 | -0.191 | -1.726 | 0.4 | -5.8 | 454.9 | 47.5 |
| 3/31/2015 | 28807 | 0.2 | 148128 | 0 | -0.191 | -1.726 | 0.5 | -6.7 | 479.3 | 49.1 |
| 4/30/2015 | 28851 | 0.1 | 148511 | 0.2 | -0.191 | -1.726 | 0.3 | -6.3 | 479.3 | 51.3 |
| 5/31/2015 | 29118 | 0.2 | 148817 | 0.1 | -0.191 | -1.726 | 0.5 | -6 | 479.3 | 49.7 |
| 6/30/2015 | 29132 | 0.2 | 148816 | 0.2 | -0.191 | -1.726 | 0.3 | -5.8 | 430.5 | 48.1 |
| 7/31/2015 | 29150 | 0.1 | 148830 | 0 | -0.191 | -1.726 | 0.5 | -6.2 | 430.5 | 48.1 |
| 8/31/2015 | 29147 | 0 | 149181 | 0.1 | -0.191 | -1.726 | 0.3 | -7.1 | 430.5 | 46.9 |
| 9/30/2015 | 29281 | -0.1 | 148826 | -0.1 | -0.191 | -1.726 | -0.1 | -7.3 | 445.7 | 42 |
| 10/31/2015 | 29298 | 0 | 149246 | -0.2 | -0.191 | -1.726 | 0.1 | -6.7 | 445.7 | 47.3 |
| 11/30/2015 | 29342 | 0.1 | 149463 | 0.1 | -0.191 | -1.726 | 0.2 | -6.5 | 445.7 | 45.5 |
| 12/31/2015 | 10000 | -0.1 | 150128 | 0.1 | -0.191 | -1.726 | 0.2 | -6.6 | 449.5 | 47.2 |
| 1/31/2016 | 10000 | 0 | 150653 | 0.4 | -0.191 | -1.726 | 0.2 | -5.9 | 449.5 | 47.3 |
| 2/29/2016 | 10000 | -0.1 | 150939 | 0 | -0.191 | -1.726 | 0.6 | -6.1 | 449.5 | 47.8 |
| 3/31/2016 | 10000 | 0.2 | 151218 | 0.1 | -0.191 | -1.726 | -0.1 | -6.2 | 479.7 | 46.8 |
| 4/30/2016 | 10000 | 0.3 | 151074 | 0.2 | -0.191 | -1.726 | 0.6 | -5.1 | 479.7 | 46.3 |
| 5/31/2016 | 10000 | 0.2 | 151132 | 0 | -0.191 | -1.726 | 0.3 | -4.5 | 479.7 | 48.7 |
| 6/30/2016 | 10000 | 0.2 | 151223 | 0.2 | -0.191 | -1.726 | 0.6 | -3.5 | 418 | 48.2 |
| 7/31/2016 | 10000 | 0 | 151554 | 0.2 | -0.191 | -1.726 | 0.2 | -3 | 418 | 45.5 |
| 8/31/2016 | 10000 | 0.1 | 151779 | 0.1 | -0.191 | -1.726 | 0.2 | -2.4 | 418 | 48.4 |
| 9/30/2016 | 10000 | 0.2 | 151761 | 0.1 | -0.191 | -1.726 | 0.5 | -1.5 | 406.7 | 46.7 |
| 10/31/2016 | 10000 | 0.2 | 151793 | 0.1 | -0.191 | -1.726 | 0.2 | -1.1 | 406.7 | 51.3 |
| 11/30/2016 | 10000 | 0 | 151954 | 0.2 | -0.191 | -1.726 | 0.2 | -0.2 | 406.7 | 51.4 |
| 12/31/2016 | 10000 | 0.2 | 152157 | 0.2 | 0.317 | -1.726 | 0.9 | 1.3 | 404.5 | 54.8 |
| 1/31/2017 | 10000 | 0.4 | 152152 | 0.3 | -0.242 | -1.726 | 0.4 | 2.4 | 404.5 | 55.6 |
| 2/28/2017 | 10000 | 0.1 | 152480 | 0.1 | 0.189 | -1.726 | 0.2 | 3.2 | 404.5 | 56.4 |
| 3/31/2017 | 10000 | -0.1 | 153065 | 0.1 | 0.481 | -1.726 | 0.4 | 3.4 | 409.1 | 55.3 |
| 4/30/2017 | 10000 | 0.2 | 153255 | 0.5 | -0.046 | -1.726 | 0.2 | 3.1 | 409.1 | 51.7 |
| 5/31/2017 | 10000 | 0 | 153069 | 0.1 | 0.136 | -1.726 | 0 | 1.4 | 409.1 | 51.3 |
| 6/30/2017 | 10000 | 0.1 | 153318 | 0.2 | 0.287 | -5.665 | 0.4 | 0.6 | 432.7 | 51.3 |
| 7/31/2017 | 10000 | 0 | 153569 | 0.1 | 0.142 | -4.827 | 0.3 | 0.9 | 432.7 | 50.2 |
| 8/31/2017 | 10000 | 0.2 | 153503 | 0.1 | 0.028 | 0.476 | 0.2 | 2.3 | 432.7 | 52.2 |
| 9/30/2017 | 10000 | 0.4 | 154286 | 0.2 | -0.075 | 4.91 | 0.9 | 2.8 | 447.7 | 53.4 |
| 10/31/2017 | 10000 | 0.1 | 153609 | 0.3 | 0.063 | -7.697 | 0.2 | 2.7 | 447.7 | 50.3 |
| 11/30/2017 | 10000 | 0.1 | 153805 | 0.3 | 0.264 | -1.529 | 0.7 | 3.1 | 447.7 | 53.6 |

| Date | | | | | | | | | |
|---|---|---|---|---|---|---|---|---|---|
| 12/31/2017 | 10000 | 0.1 | 153904 | 0.1 | 0.309 | 1.827 | 0.9 | 2.8 | 444.2 | 51.9 |
| 1/31/2018 | 10000 | 0.4 | 154425 | 0.5 | -0.826 | 2.692 | 0.3 | 3.3 | 444.2 | 55.1 |
| 2/28/2018 | 7500 | 0.2 | 155197 | 0.3 | -0.209 | 1.914 | 0.3 | 3.1 | 444.2 | 56.7 |
| 3/31/2018 | 7500 | 0.1 | 155214 | 0.4 | 0.024 | 1.123 | 0.4 | 3.4 | 270.2 | 55.6 |
| 4/30/2018 | 7500 | 0.2 | 155312 | 0.1 | -0.128 | 1.602 | 0.3 | 3.7 | 270.2 | 52.6 |
| 5/31/2018 | 7500 | 0.2 | 155652 | 0.2 | 0.4 | -1.682 | 0.5 | 5 | 270.2 | 53.6 |
| 6/30/2018 | 6825 | 0.1 | 155762 | 0.3 | -0.149 | -3.069 | 0.2 | 5.3 | 280 | 53.9 |
| 7/31/2018 | 6825 | 0.1 | 156146 | 0.4 | 0.124 | -0.791 | 0.3 | 4.3 | 280 | 56.4 |
| 8/31/2018 | 6825 | 0.1 | 155504 | 0.1 | 0.198 | 1.004 | 0.3 | 3.6 | 280 | 58 |
| 9/30/2018 | 6825 | 0.2 | 156015 | 0.3 | -0.158 | -1.703 | 0 | 2.7 | 276.5 | 55.7 |
| 10/31/2018 | 6825 | 0.2 | 156391 | 0.3 | -0.11 | -3.459 | 0.5 | 3.1 | 276.5 | 57.8 |
| 11/30/2018 | 6825 | 0 | 156721 | 0.2 | 0.127 | 3.611 | 0.6 | 1.8 | 276.5 | 56.4 |
| 12/31/2018 | 6825 | 0.1 | 156817 | 0 | 0.045 | 5.492 | -0.8 | 1.1 | 281.4 | 52.6 |
| 1/31/2019 | 6825 | -0.1 | 156487 | 0.2 | -0.468 | 2.15 | 0.2 | -0.2 | 281.4 | 52.3 |
| 2/28/2019 | 6825 | 0.2 | 156863 | 0.1 | -0.235 | -3.883 | 0.2 | 0.2 | 281.4 | 50.3 |
| 3/31/2019 | 6825 | 0.2 | 156701 | 0.2 | 0.188 | -2.59 | 0.9 | 0.6 | 303.4 | 55.7 |
| 4/30/2019 | 6825 | 0.3 | 156625 | 0.3 | 0.125 | -1.1 | 0.3 | 0.2 | 303.4 | 54.2 |
| 5/31/2019 | 6825 | 0.1 | 156821 | 0.2 | -0.096 | 2.696 | 0.4 | -0.9 | 303.4 | 58.6 |
| 6/30/2019 | 6825 | 0 | 157232 | 0 | 0.305 | -0.837 | 0.4 | -1.6 | 312 | 53.2 |
| 7/31/2019 | 6825 | 0.1 | 157529 | 0.2 | 0.067 | -3.135 | 0.6 | -0.9 | 312 | 56.6 |
| 8/31/2019 | 6825 | 0 | 157829 | 0.1 | -0.306 | -0.877 | 0.4 | -1.4 | 312 | 55.1 |
| 9/30/2019 | 6825 | 0.1 | 158339 | 0 | -0.486 | 2.133 | 0.2 | -1.7 | 322.8 | 50.8 |
| 10/31/2019 | 6825 | 0.2 | 158545 | 0 | 0.221 | -4.702 | 0.2 | -2.2 | 322.8 | 52.6 |
| 11/30/2019 | 6825 | 0.1 | 158629 | 0.1 | 0.376 | 2.931 | 0.6 | -1.3 | 322.8 | 52.9 |
| 12/31/2019 | 6825 | 0.3 | 158845 | 0.2 | 0.074 | -0.776 | 0.4 | -0.9 | 311.6 | 57 |
| 1/31/2020 | 6825 | 0.1 | 158486 | 0.4 | -0.136 | 1.963 | 0.5 | 0.4 | 311.6 | 57.4 |
| 2/29/2020 | 6825 | 0.1 | 158683 | -0.2 | -0.067 | -3.89 | -0.1 | -1.4 | 311.6 | 59.8 |
| 3/31/2020 | 6825 | -0.3 | 155371 | -0.2 | -0.928 | 6.085 | -6.9 | -3.5 | 322.8 | 53.9 |
| 4/30/2020 | 6825 | -0.4 | 133185 | -0.8 | 0.011 | -1.108 | -11.4 | -7 | 322.8 | 47.8 |
| 5/31/2020 | 6825 | 0.1 | 137133 | 0.1 | -0.05 | 1.706 | 8.3 | -6.7 | 322.8 | 49.7 |
| 6/30/2020 | 6825 | 0.3 | 142345 | 0.3 | 0.297 | -2.459 | 6 | -4.5 | 301.7 | 47 |
| 7/31/2020 | 6825 | 0.3 | 143787 | 0.4 | 0.153 | -1.748 | 1.9 | -3.7 | 301.7 | 44 |
| 8/31/2020 | 6825 | 0.3 | 147261 | 0.2 | -0.798 | -5.761 | 1.1 | -2.7 | 301.7 | 46.8 |
| 9/30/2020 | 6825 | 0.2 | 147674 | 0.4 | 0.374 | 1.623 | 1.5 | -1.8 | 299 | 45 |
| 10/31/2020 | 6825 | 0.1 | 149949 | 0.3 | 0.244 | 2.5 | 0.3 | -1.7 | 299 | 55.2 |
| 11/30/2020 | 6825 | 0.1 | 149914 | 0.2 | 0.091 | 3.402 | -0.1 | -1 | 299 | 50 |
| 12/31/2020 | 6825 | 0.4 | 149989 | 0.3 | 0.164 | -0.718 | 0.8 | 0.4 | 294.5 | 49 |